\documentclass[journal,twoside,web]{ieeecolor}
\usepackage{generic}
\usepackage{cite}
\usepackage{amsmath,amssymb,amsfonts}
\usepackage{algorithmic}
\usepackage{graphicx}
\usepackage{textcomp}
\usepackage{color,xcolor}
\usepackage{array}
\usepackage{float}
\usepackage{url}
\usepackage{threeparttable}
\usepackage{marvosym}
\usepackage{makecell}
\usepackage{amsmath}
\usepackage{array}
\usepackage{makecell}
\usepackage{booktabs}
\usepackage{amssymb}
\usepackage{multirow}
\usepackage{footnote}
\usepackage{todonotes}
\def\BibTeX{{\rm B\kern-.05em{\sc i\kern-.025em b}\kern-.08em
		T\kern-.1667em\lower.7ex\hbox{E}\kern-.125emX}}

\begin{document}
	\title{Reconstruction-driven Dynamic Refinement based Unsupervised Domain Adaptation for Joint Optic Disc and Cup Segmentation}
	\author{
		Ziyang Chen, Yongsheng Pan,~\IEEEmembership{Member,~IEEE}, and Yong Xia,~\IEEEmembership{Member,~IEEE}
		\thanks{This work was supported in part by National Natural Science Foundation of China under Grant 62171377, in part by China Postdoctoral Science Foundation under Grant BX2021333 and 2021M703340, in part by the Ningbo Clinical Research Center for Medical Imaging under Grant 2021L003 (Open Project 2022LYKFZD06), and in part by the Natural Science Foundation of Ningbo City, China, under Grant 2021J052.
			({\em Z. Chen and Y. Pan contributed equally to this work. Corresponding author: Y. Xia})}
		\thanks{Z. Chen, Y. Xia are with the National Engineering Laboratory for Integrated Aero-Space-Ground-Ocean Big Data Application Technology, 
			School of Computer Science and Engineering, Northwestern Polytechnical University, Xi'an 710072 China. Y. Xia is also with the Ningbo Institute of Northwestern Polytechnical University, Ningbo 315048, China
			(e-mail:zychen@mail.nwpu.edu.cn; yxia@nwpu.edu.cn).}
		\thanks{Y. Pan is with the School of Biomedical and Engineering, ShanghaiTech University, Shanghai 201210, China (e-mail:yspan@mail.nwpu.edu.cn).}}
	
	\maketitle
	\begin{abstract}
		
		Glaucoma is one of the leading causes of irreversible blindness. Segmentation of optic disc (OD) and optic cup (OC) on fundus images is a crucial step in glaucoma screening. Although many deep learning models have been constructed for this task, it remains challenging to train an OD/OC segmentation model that could be deployed successfully to different healthcare centers. The difficulties mainly comes from the domain shift issue, \emph{i.e.}, the fundus images collected at these centers usually vary greatly in the tone, contrast, and brightness. 
		To address this issue, in this paper, we propose a novel unsupervised domain adaptation (UDA) method called \textbf{R}econstruction-driven \textbf{D}ynamic \textbf{R}efinement \textbf{Net}work (RDR-Net), where we employ a due-path segmentation backbone for simultaneous edge detection and region prediction and design three modules to alleviate the domain gap.
		The reconstruction alignment (RA) module uses a variational auto-encoder (VAE) to reconstruct the input image and thus boosts the image representation ability of the network in a self-supervised way. It also uses a style-consistency constraint to force the network to retain more domain-invariant information.
		The low-level feature refinement (LFR) module employs input-specific dynamic convolutions to suppress the domain-variant information in the obtained low-level features.
		The prediction-map alignment (PMA) module elaborates the entropy-driven adversarial learning to encourage the network to generate source-like boundaries and regions.
		We evaluated our RDR-Net against state-of-the-art solutions on four public fundus image datasets. Our results indicate that RDR-Net is superior to competing models in both segmentation performance and generalization ability.
		
	\end{abstract}
	

	\section{Introduction}
	\label{sec:introduction}
	\IEEEPARstart{G}{laucoma} 
	is a leading cause of irreversible blindness in the world, and therefore is regarded as a growing global health concern.
	Early screening for glaucoma plays an essential role in timely treatment.
	In glaucoma screening, the segmentation of the optic disc (OD) and optic cup (OC) on fundus images is a crucial step, since the ratio of vertical cup diameter to vertical disc diameter, known as the cup-disc-ratio (CDR), is an important indicator used by ophthalmologists for the optic nerve head evaluation~\cite{article3}.
	
\begin{figure}[!htb]
		\centering
		\includegraphics[width=\columnwidth]{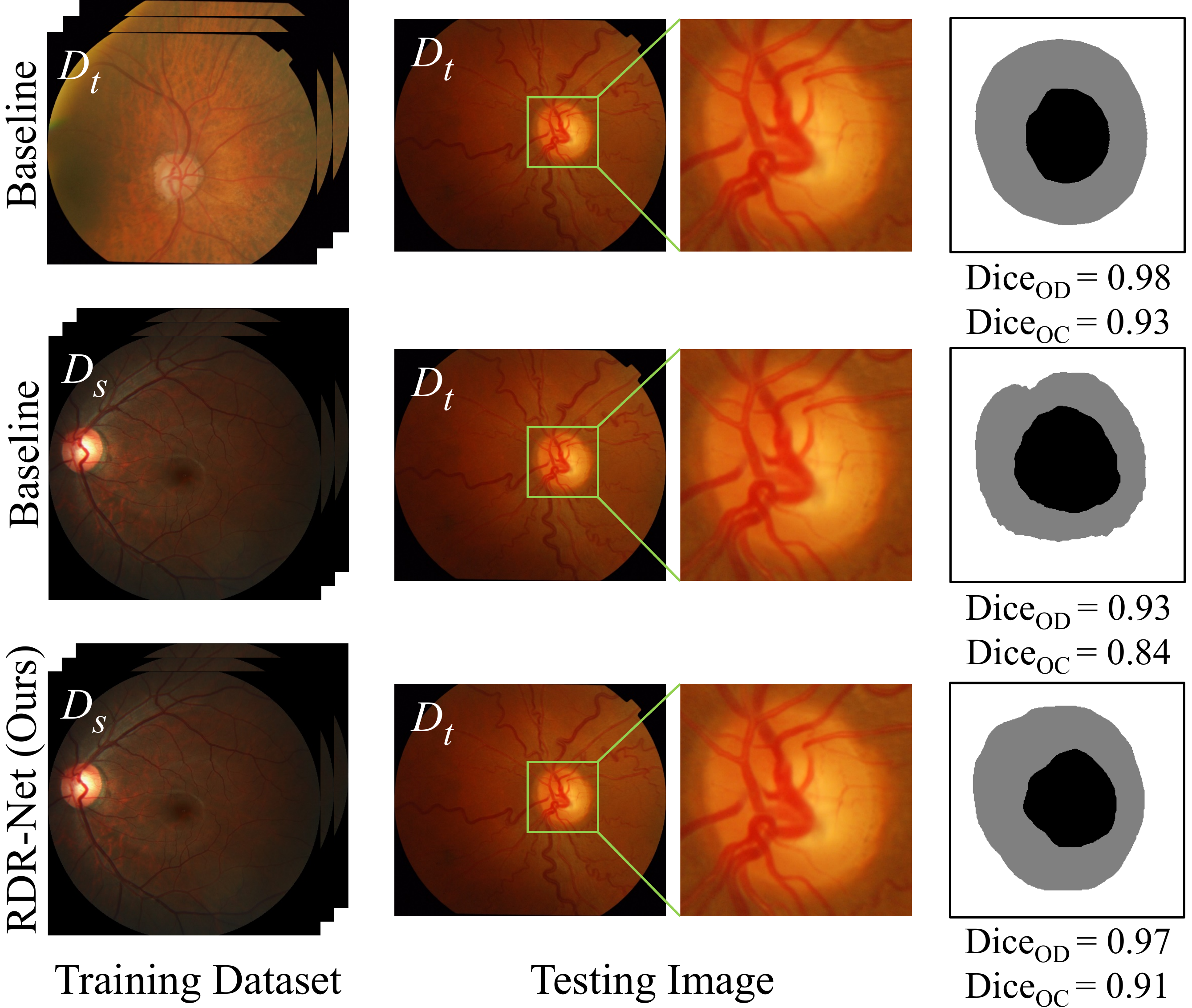}
		\caption{Illustration of domain shift issue for OD/OC segmentation. The source domain (REFUGE) and target domain (Drishti-GS) are denoted by $D_{s}$ and $D_{t}$, respectively. The segmentation network trained on $D_{s}$ performs worse on the target domain image (middle row) than the one trained on $D_{t}$ (top row), with a decrease of the Dice coefficient from $0.98$/$0.93$ to $0.93$/$0.84$ for OC (black) / OD (gray). The proposed RDR-Net is also trained on $D_{s}$, but is able to achieve substantially improved performance on the target domain image (bottom row), boosting the Dice coefficient from $0.93$/$0.84$ to $0.97$/$0.91$.}
		\label{fig3}
	\end{figure}

	To bypass the time-consuming, laborious, and highly subjective manual segmentation, automated OD/OC segmentation has been extensively studied.
	Traditionally, this task is performed by extracting manually-designed features followed by pixel classification~\cite{article5,article19,article20}. These traditional approaches, however, usually have limited performance, largely due to the insufficient representation ability of manual features.
	Recent years have witnessed the application of deep learning models to OD/OC segmentation~\cite{article6,article7,article8,article9,article71}, aiming to address the difficulties such as insufficient training samples and low target-background contrast.
	Unfortunately, there remains a major hurdle on the path between training an OD/OC segmentation model in the lab and applying it to clinical practices. This is the domain shift issue caused by the variations among multiple image domains. The fundus images collected at different healthcare centers usually vary greatly in the tone, contrast, and brightness and this relates to the diversity in imaging instruments, lighting conditions, operators, and patients.
	Due to this issue, the representations learned by a segmentation network on the source domain can hardly be applied to the target domain effectively, resulting in worse performance than the one trained on the same (target) domain (see the top and middle rows in Fig~\ref{fig3}).

	A trivial solution to domain shift is to train the network on both source and target domains, which requires the annotations of target domain data for training and is difficult to popularize. Alternatively, the unsupervised domain adaptation (UDA) emerges to be a promising paradigm that is able to achieve competitive performance without using target domain annotations~\cite{article10,article11,article12,article13,article22,article23,article24,article25}.
	To alleviate the domain shift, many UDA methods use adversarial learning to perform feature alignment, aiming to enable the segmentation network to focus on domain-invariant features. However, due to the lack of target domain annotations, the unsupervised representation learning on the target domain cannot provide precise guidance to the network and leads to a poor representation ability.
	Inspired by the image reconstruction-based self-supervision methods~\cite{article61,article62}, we introduce a reconstruction branch to the network, forcing the network to learn the representations on the target domain in this self-supervised way.
	
	Meanwhile, the low-level features extracted by the shallow layers of a network are also seriously affected by the domain shift~\cite{article60,article66}, due to the existence of both domain-invariant and domain-variant information in these features. However, we shall not neglect these features since they are beneficial to semantic segmentation~\cite{article58,article59}.
	Since high-level features are commonly recognized to contain more abstract and semantic information and are less domain-variant~\cite{article63}, we advocate using the high-level features extracted by deep layers to refine low-level features for improved robustness to the domain shift.
	
	Moreover, the low-level feature refinement should ideally be conducted in an input-specific way, since low-level features depends largely on the input image. Unfortunately, mainstream feature alignment methods~\cite{article12,article13} are based on traditional convolutions and not competent for this task, due to their frozen parameters in the inference phase.
	The recent advances in dynamic convolutions~\cite{article21,article14,article15,article16} enable a neural network to adapt its parameters to different input samples in the inference phase, and thus boost the generalization of the network.
	For example, Zhang \emph{et al.}~\cite{article16} presented a dynamic on-demand segmentation head, in which convolutional parameters are determined by a controller on condition of the task coder and the features of input image. Thus the network can adaptively segment different organs and tumors and the parameters in the dynamic head can be adjusted in the inference stage.
	Prompted by such successful applications, we argue that the dynamic convolution could be a promising tool for our input-specific low-level feature refinement.

	In this paper, we propose a novel UDA method called \textbf{Reconstruction}-driven \textbf{D}ynamic \textbf{R}efinement \textbf{Net}work (RDR-Net) to overcome the domain shift issue for joint OD/OC segmentation on fundus images.
	RDR-Net is composed of a due-path segmentation backbone, a reconstruction alignment (RA) module, a low-level feature refinement (LFR) module, and a prediction-map alignment (PMA) module.
	The segmentation backbone has one encoder and two decoders. The encoder is used to extract low-level features (by shallow layers) and high-level features (by deep layers). While the decoders are used for edge detection and region prediction, respectively.
	The RA module uses a variational auto-encoder (VAE)~\cite{article17} to reconstruct the input image and thus boosts the image representation ability of the network in a self-supervised way. Moreover, a style-consistency constraint is added to the RA module to force the network to retain more structure information, which is domain-invariant.
	The LFR module uses input-specific dynamic convolution to suppress the domain-variant information in the obtained low-level features. The parameters in those convolutions are generated based on the data distribution estimated by the RA module and the structure information in the high-level features.
	We also adopt the entropy-driven adversarial learning~\cite{article12} to construct the PMA module, which is used to further encourage the network to generate source-like boundaries and region predictions.

	The uniqueness of our RDR-Net is that it addresses the domain shift issue from three aspects simultaneously. 
	First, it employs VAE to perform input image reconstruction, enabling the model to learn image representations on the target domain in a self-supervised way.
	Second, it uses dynamic convolutions to suppress the domain-variant information in low-level features.
	Third, it adopts the adversarial learning to align the boundaries and regions obtained on the source domain and target domain.
	We evaluated the proposed RDR-Net against several state-of-the-art methods on four public fundus image datasets. Our results suggest the effectiveness of each proposed module, and also indicate that the OD/OC segmentation performance and generalization ability of RDR-Net are superior to those of competing methods.

	\section{Related Work}
	\label{sec:RelatedWorks}
	
	\subsection{Joint OD/OC Segmentation}
	Joint OD/OC segmentation on fundus images has been thoroughly studied. 
	Fu \emph{et al.}~\cite{article6} employed the image pyramid input to extract multi-scale features and utilized the polar transformation to balance the proportion between OD and OC.
	Liu \emph{et al.}~\cite{article8} adopted depthwise separable convolutional layers to construct the dense depthwise separable convolutional block for improved segmentation accuracy.
	Besides these pixel-wise dense prediction methods, there are combined solutions in which the region proposal network (RPN) is incorporated into the segmentation framework~\cite{article9,article7}.
	Jiang \emph{et al.}~\cite{article9} designed the OD PRN and OC RPN for localization and employed the attention mechanism to guide the localization of OC.
	Yin \emph{et al.}~\cite{article7} proposed a segmentation based RPN and a pyramid RoI Align module to improve the accuracy of proposals and aggregate the multi-level information.
	Despite these solutions, the accurate segmentation of OD and OC, particularly OC, remains a challenging task, since the OD-OC contrast is low and the edge of OC is blurry.
	In this work, we employed the edge adversarial learning to use the edge information for accurate segmentation and explored UDA for better generalization on test data.

	\subsection{UDA Methods}
	Domain adaptation aims to refine the deep network when faced with the distribution shift between source (training) and target (test) domains~\cite{article26,article27}.
	Although several metrics, such as the maximum mean discrepancy (MMD) with various kernels~\cite{article25}, have been proposed to characterize the domain discrepancy, these metrics usually suffer from limited expressiveness~\cite{article24}.
	Recently, UDA methods, which do not need the manual annotations of target domain samples for training, have drawn increasing research attention, particularly in the field of medical image analysis~\cite{article10,article11,article12,article22,article23,article13,article28,article35,article36,article52}.
	The most commonly used strategies for UDA can be roughly categorized into feature alignment and image synthesis.
	
	It is acknowledged that there are domain-invariant features such as the shape and structure of the regions of interest (RoIs), which do not vary a lot across domains. 
	The first group of UDA methods are based on the idea of using adversarial learning to minimize the distribution discrepancy between the features or segmentation results obtained by different domains.
	Hoffman \emph{et al.}~\cite{article52} presented the first UDA method for semantic image segmentation, which combines global and category specific adaptation using adversarial training.
	Kamnitsas \emph{et al.}~\cite{article22} proposed a multi-connected domain discriminator for improved adversarial learning and forced the segmentation network to extract domain-invariant features by adversarial training.
	Javanmardi \emph{et al.}~\cite{article23} utilized a domain classifier in an adversarial setting to learn a cross-domain loss and thus alleviated the domain shift.
	Wang \emph{et al.}~\cite{article11} designed an effective morphology-aware segmentation loss and a patch-based discriminator to obtain local structure information.
	Wang \emph{et al.}~\cite{article12} incorporated adversarial learning into the network to produce low-entropy and stable predictions. 
	Zhang \emph{et al.}~\cite{article10} introduced the attention mechanism and adopted adversarial learning of attention maps and feature maps to locate and extract domain-invariant features across different datasets.
	However, these methods overlook the low-level features, which are important for the semantic segmentation task. Refining low-level features can enhance the final performance~\cite{article64}.
	
	Since generative adversarial networks (GANs)~\cite{article34,article28} can perform cross-domain image translation based on unpaired images, image synthesis has been widely used in UDA to reduce the domain gap via mapping the samples from different domains into an intermediate latent space.
	Huo \emph{et al.}~\cite{article35} designed an end-to-end segmentation network, which can perform cross-modality image synthesis and supervised image segmentation.
	Kamnitsas \emph{et al.}~\cite{article22} incorporated the edge structure into cycle-consistency GAN (CycleGAN) to generate high-quality images for domain adaptation.
	Zhang \emph{et al.}~\cite{article36} proposed another variant of CycleGAN for pixel-level translation and utilized a pre-trained module to enforce the segmentation consistency between different domains.
	Lei \emph{et al.}~\cite{article13} utilized a variant of CycleGAN to generate target-like query images, and adopted both style-consistency constraint and content consistency constraint to alleviate the domain shift.
	Although these methods have achieved performance gains, they still suffer from the high computational complexity and troublesome training of the generative network.
	
	Similar to these methods, our RDR-Net also adopts adversarial 
	learning for the feature-level alignment. However, to reduce the complexity and difficulty of training, we replace CycleGAN with a VAE branch for image synthesis. Moreover, existing methods may overlook the low-level features and do not learn the image representation in a reconstruction-based way. By contrast, our RDR-Net uses VAE for representation learning and constructs dynamic convolutions to refine the low-level features. Consequently, our RDR-Net can be trained in an end-to-end manner and the encoded latent variable obtained by VAE can be used as a heuristic to guide the segmentation network.

	\subsection{Application of Dynamic Convolution}
	Dynamic convolution is able to adaptively adjust its convolutional parameters according to the input image, and hence is a far more flexible operation with strong self-adaptability than its static counterpart~\cite{article14,article15,article16,article21}.
	Jia \emph{et al.}~\cite{article21} developed a dynamic filter network to increase the flexibility of network by generating convolutional kernels dynamically conditioned on the input image. 
	Yang \emph{et al.}~\cite{article14} designed a conditionally parameterized network called CondConv, whose inference capacity is increased by learning specialized convolutional kernels for each input image.
	Chen \emph{et al}~\cite{article15} proposed the dynamic convolution, which is a linear mixture of multiple convolutional layers whose parameters are calculated according to the input-related attention.
	Generally, dynamic convolutions have distinct advantages over traditional ones, such as improving the network flexibility without excessively increasing parameters~\cite{article21}, incorporating the attention mechanism into convolutional kernels~\cite{article15}, and increasing the generalization and adaptation to an assigned task~\cite{article14,article16}. However, dynamic convolutions are mostly designed to adapt the network to an input image, instead of a new domain. In our RDR-Net, we employ VAE to obtain the data distributions of input images, which is then used to generate dynamic filters based on the corresponding domain for domain adaptation.
	
	\section{Methodology}
	\label{sec:Methodology}
	
\begin{figure*}[!htb]
		\centering
		\includegraphics[width=\textwidth]{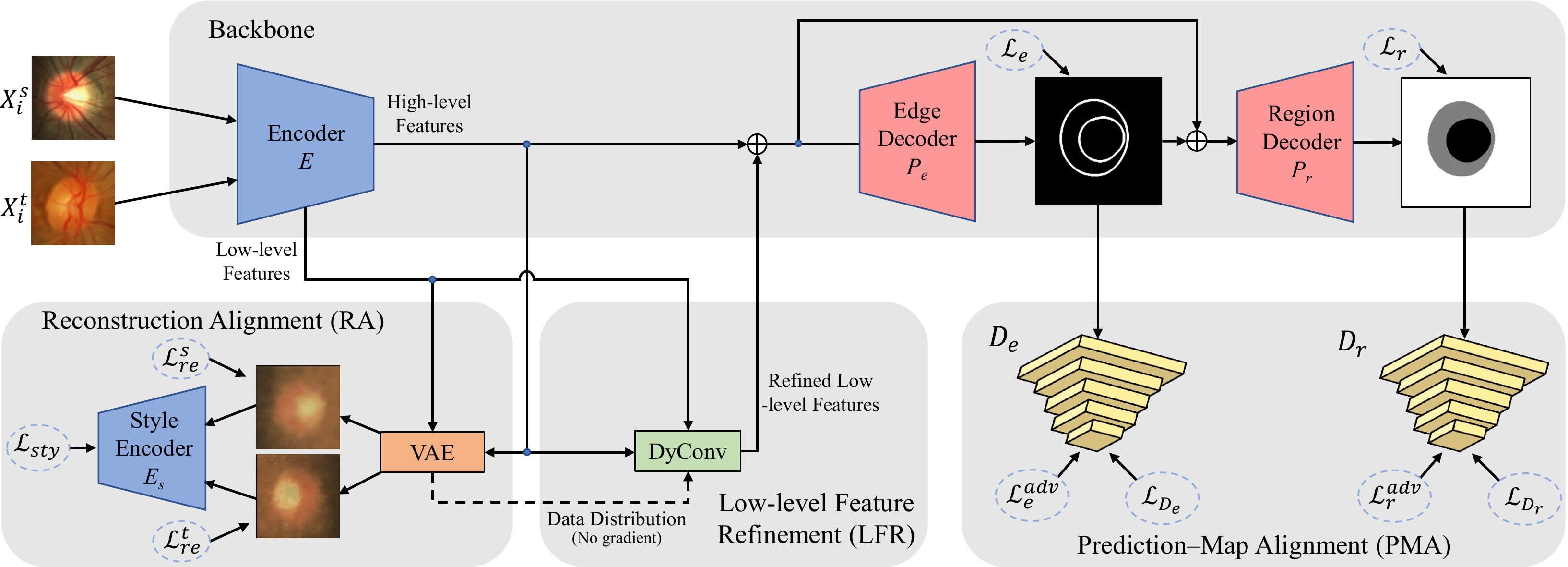}
		\caption{Overview of the proposed RDR-Net, which consists of a backbone network, an RA module, a LFR module, and a PMA module. 
			The backbone uses an encoder for feature extraction and two decoders for edge detection and region prediction, respectively. It is trained on source domain by minimizing the edge loss $\mathcal{L}_{e}$ and region prediction loss $\mathcal{L}_{r}$.
			The RA module uses VAE for image reconstruction via minimizing the reconstruction loss $\mathcal{L}_{re}^{s}$ and $\mathcal{L}_{re}^{t}$. This module also contains a style encoder $E_{s}$ with a style-consistency loss $\mathcal{L}_{sty}$ for features alignment.
			The LFR module uses a dynamic convolution (DyConv) block to refine the low-level features. The parameters in DyConv are generated based on the distribution information given by VAE and the semantic information from high-level features.
			The PMA module uses two discriminators ($D_{e}$ and $D_{r}$), which are trained by optimizing the cross-entropy loss ($\mathcal{L}_{D_{r}}$ and $\mathcal{L}_{D_{e}}$), to align the distributions of edge predictions and region predictions via minimizing the adversarial loss ($\mathcal{L}_{e}^{adv}$ and $\mathcal{L}_{r}^{adv}$).
			The dotted line indicates that the gradient will not propagate back, and $\oplus$ denotes the concatenate operation.}
		\label{fig4}
	\end{figure*}

	Let a set of source domain images be denoted by 
	$\mathcal{D}^{s} = \left \{ X_{i}^{s}, Y_{i}^{s} \right \}_{i=1}^{N^s}$, where $X_{i}^{s}\in\mathbb{N}^{H \times W \times 3}$ is an image and $Y_{i}^{s}\in \mathbb{R}^{H \times W}$ is the corresponding segmentation ground truth.
	The edge map of $X_{i}^{s}$, denoted by $B_{i}^{s}\in \mathbb{R}^{H \times W}$, is obtained by applying the Sobel operator and Gaussian filter to $Y_{i}^{s}$.
	Similarly, a set of unlabeled target domain images is denoted by $\mathcal{D}^{t} = \left \{ X_{i}^{t} \right \}_{i=1}^{N^t}$. 
	The proposed RDR-Net is trained on $\left \{ \mathcal{D}^{s},\mathcal{D}^{t} \right \}$ and tested on $\mathcal{D}^{t}$.
	RDR-Net consists of a two-decoder segmentation backbone, a RA module, an LFR module, and a PMA module. Feeding an image $X$ from either $\mathcal{D}^s$ or $\mathcal{D}^t$ into the encoder, we obtain the low-level features $F_l$ and high-level features $F_h$. Then, we feed both $F_l$ and $F_h$ into the VAE branch to estimate the data distribution $z$ and perform image reconstruction. The reconstructed image $R$ is not only used to calculate the reconstruction loss using Eq.~(\ref{eq-lre}), but also fed into a style encoder to extract its style features, which is used to calculate the style-consistency loss using Eq.~(\ref{eq-lsty}). Next, the data distribution $z$ and high-level features $F_h$ are processed and utilized to generate the parameters of the dynamic convolutional layers. The dynamic convolutional layers convert $F_l$ into the refined low-level features $F_r$, which are concatenated with $F_h$ to form the fusion feature $F_s$. We feed $F_s$ into the edge decoder to predict an edge map, and the concatenation of $F_s$ and the edge map is fed into the region decoder to predict a region map. The supervised losses shown in Eq.~(\ref{eq-lseg}) and (\ref{eq-ledge}) are used to optimize the network. We also construct two discriminators and use the adversarial losses calculated by Eq.~(\ref{eq-ladv_r}) and (\ref{eq-ladv_e}) to further improve the learning process.
	The diagram of RDR-Net is shown in Fig.~\ref{fig4}. We now delve into the details of each part.

	\subsection{Backbone}
	The segmentation backbone consists of a shared encoder for feature extraction, a decoder for edge detection, and a decoder for region prediction~\cite{article12}.
	The encoder $E$ is constructed based on DeepLabv3+~\cite{article30} with a pre-trained MobileNetV2 backbone~\cite{article29} and hence has a lightweight architecture.
	The edge decoder $P_{e}$ consists of three convolutional layers with $256$, $256$, and $1$ channels, respectively~\cite{article12}.
	Each of the first two layers is followed by the ReLU activation and batch normalization, and the last layer is followed by the {\it sigmoid} activation.
	Since the edge structure can be regarded as a domain-invariant feature, the object boundaries generated by $P_{e}$ provide valuable guidance for object region prediction.
	The region decoder $P_{r}$ contains only one convolutional layer that is followed by {\it sigmoid} activation~\cite{article12}.
	
	Feeding an image $X$, which is from either $\mathcal{D}^{s}$ or $\mathcal{D}^{t}$, to the backbone, we define the output of the second bottleneck of $E$ as the low-level feature $F_{l}$, which has $24$ channels, and define the final output of $E$ as the high-level feature $F_{h}$, which has $320$ channels.
	Then, $F_{l}$ is converted into the refined low-level feature $F_{r}$ by the LFR module (in Section~\ref{LFR}).
	And $F_{h}$ is first up-sampled using the bilinear interpolation, and then concatenated with $F_{r}$ to form the fused feature $F_{s}$.
	Based on $F_{s}$, the edge decoder $P_{e}$ predicts an edge map $\hat{B}\in \mathbb{R}^{\frac{H}{4} \times \frac{W}{4}}$.
	Taking both $F_{s}$ and $\hat{B}$ as its input, the region decoder $P_{r}$ produces a predicted region map $\hat{Y}\in \mathbb{R}^{\frac{H}{4} \times \frac{W}{4}}$. Then, the $\hat{B}$ and $\hat{Y}$ are up-sampled to the size of input images using the bilinear interpolation.
	
	We train the backbone on source domain data in a supervised way. Given the predicted region map $\hat{Y}_i^s$ and corresponding ground truth $Y_i^s$, the region prediction loss is defined as
	\begin{equation}
		\begin{aligned}
			\mathcal{L}_{r}(\hat{Y}^{s}, Y^{s}) = \mathcal{L}_{ce}(\hat{Y}^{s}, Y^{s}) + \mathcal{L}_{gdl}(\hat{Y}^{s}, Y^{s}).
		\end{aligned}
		\label{eq-lseg}
	\end{equation}
	where $\mathcal{L}_{ce}$ is the cross-entropy loss, and $\mathcal{L}_{gdl}$ is the generalized Dice loss (GDL), which is calculated as~\cite{article40}
	\begin{equation}
		\begin{aligned}
			\mathcal{L}_{gdl}(\hat{Y}^{s}, Y^{s}) = 1 - 2\frac{\sum_{l=1}^{2}w_{l}\sum \hat{Y}_{l}^{s}Y_{l}^{s}}{\sum_{l=1}^{2}w_{l}\sum \left (\hat{Y}_{l}^{s}+Y_{l}^{s} \right )},
		\end{aligned}
	\end{equation}
	where $w_{l}$ denotes the weight for category $l$. We have two foreground categories, \emph{i.e.}, OC and OD. Since OC always lies within OD in normal eyes, we define the weights $\lbrace w_{1},w_{2}\rbrace$ as follows to balance the contributions of both categories
	\begin{equation}
		\begin{aligned}
			w_{l} = 1 - \frac{\sum Y_{l}^{s}}{\sum_{l=1}^{2}\sum Y_{l}^{s}}.
		\end{aligned}
	\end{equation}

	Meanwhile, although OC and OD have similar structures, the edge between them is blurred and hard to identify. To optimize the network towards producing the accurate boundaries of OC and OD, we also define the following edge loss~\cite{article12} for edge prediction
	\begin{equation}
		\begin{aligned}
			\mathcal{L}_{e}(\hat{B}^{s}, B^{s}) = \frac{1}{M}\| B^{s}-\hat{B}^{s}\|_2^{2},
		\end{aligned}
		\label{eq-ledge}
	\end{equation}
	where $M = H \times W$ is the number of pixels.

	\subsection{RA Module}
\begin{figure}[!htb]
		\centering
		\includegraphics[width=\columnwidth]{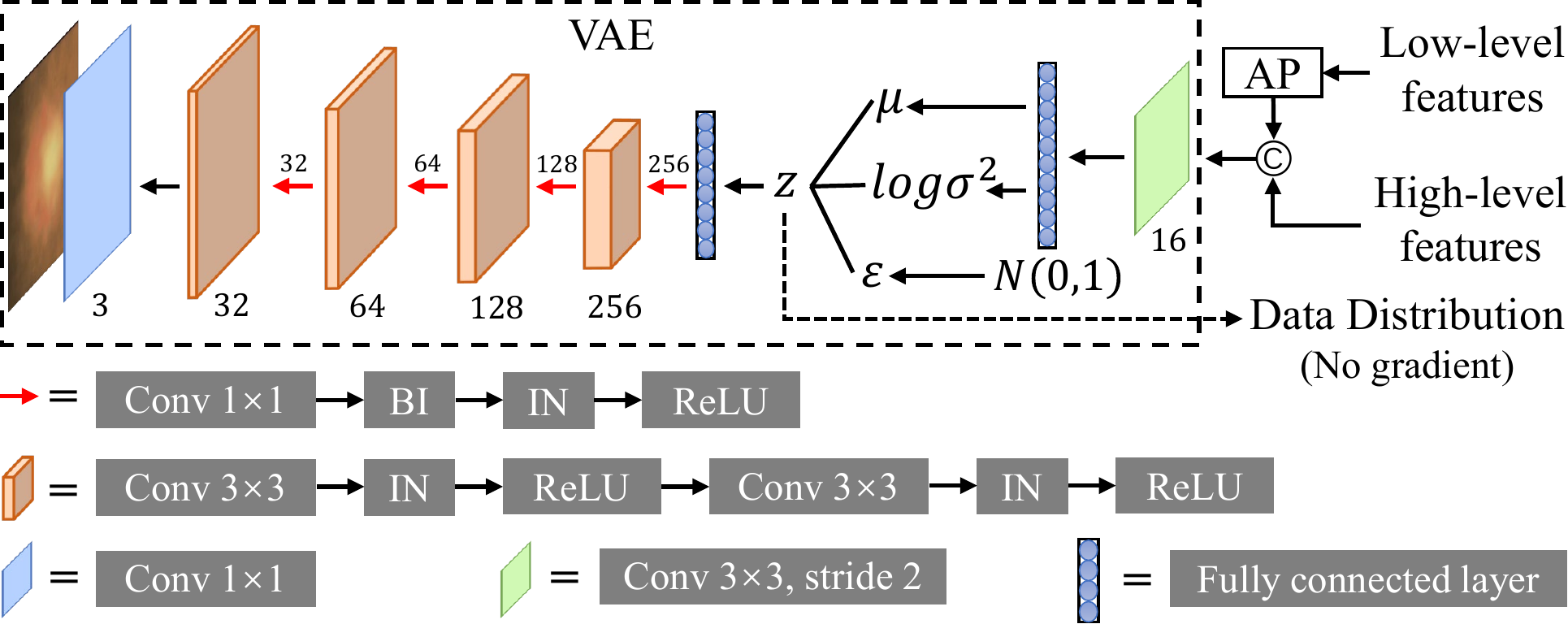}
		\caption{Diagram of VAE branch. AP: Average pooling; IN: Instance normalization; and BI: Bilinear interpolation for $2\times2$ up-sampling.}
		\label{fig5}
	\end{figure}

	In the RA module, we employ VAE~\cite{article17} (see Fig.~\ref{fig5}) to perform reconstruction and feature alignment. 
	Since VAE explores the distribution of inputs more explicitly than GANs, our VAE branch can provide the way of regularization aiming to force the network to enhance the representation ability and learn the domain invariant features.
	
	Specifically, we assume each image can be represented by a $D$-dimensional feature, which follows a component-independent Gaussian distribution $\mathcal{N}(\mu, \sigma^{2})$. We empirically set $D$ to $128$ for this study.
	For each input $X$, the low-level features $F_{l}$ are first down-sampled by an average pooling layer and then concatenated with the high-level features $F_{h}$ along the channel-axis. The concatenated features are dimension-reduced by a convolutional layer, flatten into a vector, and further transformed by a fully-connected layer to a $2D$-dimensional vector, which represents the concatenation of the mean $\mu$ and $\log\sigma^{2}$. Note that, since the variance is non-negative, we replace it with $\log\sigma^{2}$ for the simplicity of estimation.
	
	In the decoding process, we first sample a vector $\varepsilon$ from the standard Gaussian distribution $\mathcal{N}\left (0,1 \right )$. Then the latent feature vector $z$ that follows the Gaussian distribution $\mathcal{N}(\mu, \sigma^{2})$ can be obtained as~\cite{article17}
	\begin{equation}
		\begin{aligned}
			z = \mu + \sigma \odot \varepsilon,
		\end{aligned}
	\end{equation}
	where $\odot$ means element-wise multiplication. 
	Next, we feed $z$ to a fully connected layer, which is followed by the ReLU activation, and reshape the output into a feature map of size $\frac{H}{16}\times \frac{W}{16}\times 16$.
	Finally, this feature map is feed to the decoder that has four convolutional blocks and one convolutional layer followed by the {\it {\it sigmoid}} activation, resulting in a reconstructed image $R \in\mathbb{R}^{H \times W \times 3}$.

	The reconstruction loss is defined as
	\begin{equation}
		\begin{aligned}
			\mathcal{L}_{re} = \mathcal{L}_{kl} + \mathcal{L}_{mse},
		\end{aligned}
		\label{eq-lre}
	\end{equation}
	where the KL divergence error~\cite{article38, article17}
	\begin{equation}
		\mathcal{L}_{kl} = \frac{1}{D}\|\mu^{2}+\sigma^{2}-\log\sigma^{2}-1 \|_1,\\
	\end{equation}
	is a standard VAE penalty term to ensure the generation capability of the network, and the mean absolute error~\cite{article32}
	\begin{gather}
		\mathcal{L}_{mse} = \frac{1}{M}\|R-X\|_2^{2},
	\end{gather}
	is a common reconstruction loss for VAE. Herein, we denote the reconstruction losses for a source or a target image as $\mathcal{L}_{re}^{s}$ and $\mathcal{L}_{re}^{t}$, respectively.

	Moreover, we also utilize a style encoder $E_{s}$ to extract the style features of reconstructed images, and then impose a style-consistency constraint on style features, aiming to enforce the network to filter out domain-variant information without being affected by different image styles. After that, the VAE branch can provide the latent feature vector $z$, which is domain-invariant, to the LFR module to refine the low-level features.
	We use the first four convolutional layers of VGG19~\cite{article31} that has been pre-trained on ImageNet as $E_{s}$~\cite{article39}, and do not update its parameters during training. Let the style features of a reconstructed image be denoted by $F_i= (F_{i1}, F_{i2}, \cdots,F_{iC})$, where $C$ is the number of channels. The Gram matrix of $F_i$ is 
	\begin{equation}
		G_{i} = (\vec{F}_{ij}^{\top} \vec{F}_{ik})_{j,k\in \{1,\cdots,C\}},
	\end{equation}
	where $\vec{*}$ means vectoring the matrix $*$.
	Then, we define the style-consistency loss as
	\begin{equation}
		\begin{aligned}
			\mathcal{L}_{sty} = \frac{1}{4C^2M^{2}}\|\vec{G^{s}}-\vec{G^{t}}\|_2^2.
		\end{aligned}
		\label{eq-lsty}
	\end{equation}

	\subsection{LFR module}
	\label{LFR}
\begin{figure}[!htb] 
		\centering
		\includegraphics[width=\columnwidth]{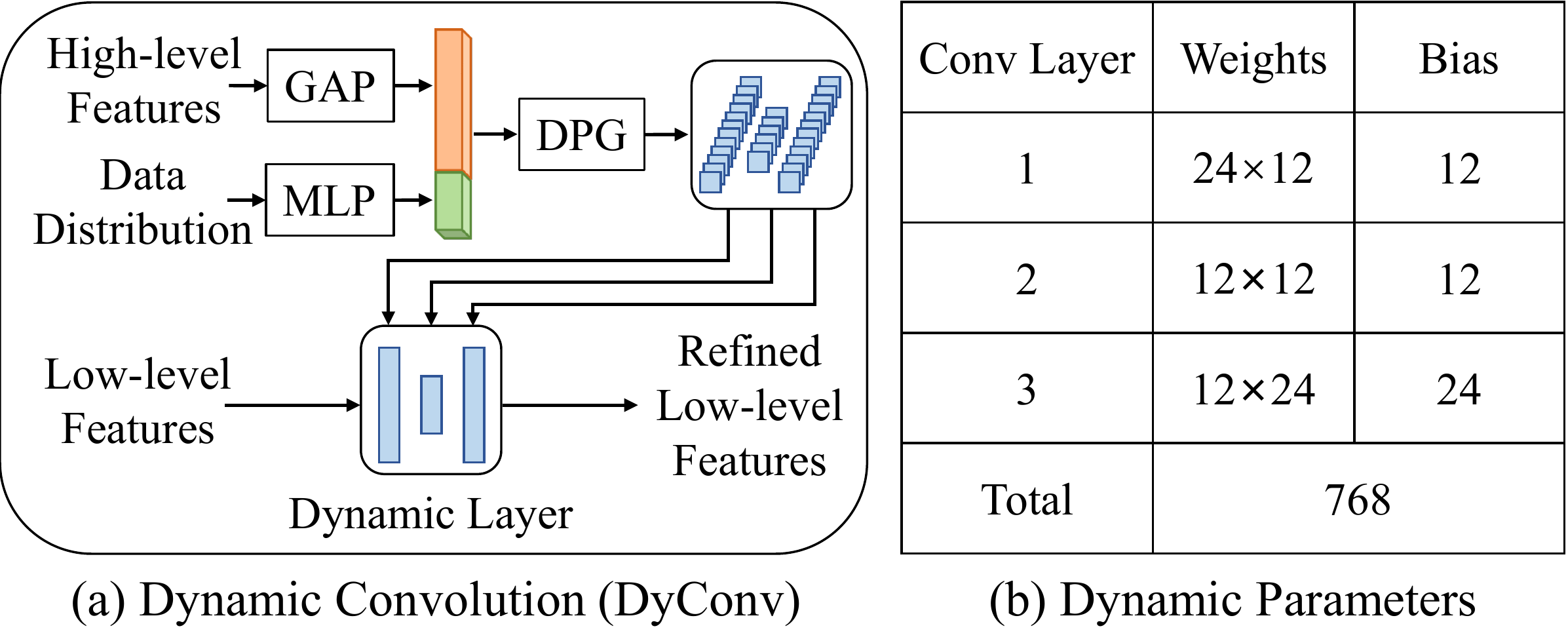}
		\caption{Illustration of DyConv block: (a) Architecture of DyConv block and (b) Number of parameters in each dynamic convolutional layer. 
			GAP: Global Average pooling; MLP: Multi-Layer Perceptron; and DPG: Dynamic Parameter Generator that is a single $1\times1$ convolutional layer.
			The parameters in three dynamic convolutional layers are conditioned on the high-level semantic information (highlighted in orange) and the distribution of input images (highlighted in green).}
		\label{fig6}
	\end{figure}

	The cross-domain generalization ability of the segmentation network depends heavily on the low-level feature $F_{l}$, which is expected to be domain-invariant.
	Therefore, we devise the dynamic convolution (DyConv) block (see Fig.~\ref{fig6}) to retain most domain-invariant information in $F_{l}$ while removing the noise.
	
	Traditional convolutional layers update learnable parameters during training but freeze them during inference. This is a main reason for the performance degradation caused by training the network on one domain and testing it on another domain. 
	To address this issue, we attempt to generate convolutional parameters dynamically and thus adapt those convolutions better to the input images from different domains.
	Specifically, we design three $1\times1$ dynamic convolutional layers with $12$, $12$, and $24$ channels, respectively, each being followed by the ReLU activation. The structure of these layers is similar to the Bottleneck block in ResNet~\cite{article49}, and the parameters in them are denoted by $\omega = \left \{ \omega_{1}, \omega_{2}, \omega_{3}\right \}$, where $\omega_{1}$ has $300$ elements, $\omega_{2}$ has $156$ elements, and $\omega_{3}$ has $312$ elements. Using these dynamic convolutional layers, the refined low-level feature $F_{r}$ can be computed as
	\begin{equation}
		\begin{aligned}
			F_{r} = \mathfrak{f} \left ( \mathfrak{f} \left ( \mathfrak{f} \left ( F_{l}\ast \omega_{1} \right )\ast \omega_{2}\right )\ast \omega_{3}\right ),
		\end{aligned}
	\end{equation}
	where $\ast$ indicates the convolutional operation, and $\mathfrak{f}$ is the ReLU activation.

	To generate $\omega$ dynamically, we utilized a traditional convolutional layer as the dynamic parameter generator.
	The input of this generator is the concatenation of two parts. One is the high-level feature $F_{h}$ pooled by GAP, and the other is the latent feature vector $z$ processed by a three-layer multi-layer perception (MLP). The MLP has 128, 64, and 64 neurons in three layers, and each of the first two layers is followed by the GeLU activation. 
	
	\subsection{PMA module}
	
	Besides supervised learning, adversarial learning is also used during training, aiming to align the domain-related features and thus encourage the network to generate source-like segmentation results on the target domain.
	
	Based on the discriminator presented in Ref.~\cite{article12}, we construct our region discriminator $D_{r}$ and edge discriminator $D_{e}$ for adversarial learning, which consists of five $4 \times 4$ convolutional layers with channels $\left [ 64, 128, 256, 512, 1 \right ]$, and each layer is followed by the Leaky ReLU activation except the last one~\cite{article42}.
	Although $D_{r}$ and $D_{e}$ have the same structure, their parameters are not shared.
	Due to the domain gap, the predictions on the target domain are prone to be uncertain and hence high-entropy~\cite{article41}.
	We employ the entropy-driven adversarial learning~\cite{article12} to suppress uncertain predictions. The region discriminator $D_{r}$ is trained to judge whether the predicted region map $\hat{Y}$ is from the source domain or target domain by minimizing the following cross-entropy loss
	\begin{equation}
		\begin{aligned}
			\mathcal{L}_{D_{r}}(\hat{Y}^{s}, \hat{Y}^t) = \mathcal{L}_{ce}\left (D_{r}\left (\mathbb{E}\left ( \hat{Y}^{s} \right )\right ), 1 \right )+ \\ \mathcal{L}_{ce}\left (D_{r}\left (\mathbb{E}\left ( \hat{Y}^{t} \right ), 0 \right )\right ),
		\end{aligned}
	\end{equation}
	where the label is set to $1$ for the source domain and $0$ for the target domain, and the entropy map $\mathbb{E}$ is calculated as
	\begin{equation}
		\begin{aligned}
			\mathbb{E}\left ( \hat{Y} \right ) = -\hat{Y} \log \left ( \hat{Y} \right ).
		\end{aligned}
	\end{equation}
	
	On the other hand, the network is encouraged to produce the source-like predicted region map on the target domain to cheat the discriminator so that it can generalize well on the target domain. To this end, we also use the following adversarial loss
	\begin{equation}
		\begin{aligned}
			\mathcal{L}_{r}^{adv}(\hat{Y}^{t}) = \mathcal{L}_{ce}\left (D_{r}\left (\mathbb{E}\left ( \hat{Y}^{t} \right )\right ), 1 \right ).
		\end{aligned}
		\label{eq-ladv_r}
	\end{equation}
	
	For the predicted edge map $\hat{B}$, we employ the edge adversarial learning~\cite{article12} and train the edge discriminator $D_{e}$ to determine whether $\hat{B}$ is from the source domain or target domain. Similarly, we have the following edge cross-entropy loss
	\begin{equation}
		\begin{aligned}
			\mathcal{L}_{D_{e}}(\hat{B}^{s}, \hat{B}^{t}) = \mathcal{L}_{ce}\left (D_{e}\left (\hat{B}^{s}\right ),1 \right )+\mathcal{L}_{ce}\left (D_{e}\left (\hat{B}^{t}\right ),0 \right ).
		\end{aligned}
	\end{equation}
	The edge adversarial loss 
	\begin{equation}
		\begin{aligned}
			\mathcal{L}_{e}^{adv}(\hat{e}^{t}) = \mathcal{L}_{ce}\left (D_{e}\left (\hat{B}^{t}\right ),1 \right ),
		\end{aligned}
		\label{eq-ladv_e}
	\end{equation}
	is used to optimize the network so as to further align the distribution of predicted edge map on different domains.
	
	In summary, the overall objective of the whole segmentation network is
	\begin{equation}
		\begin{aligned}
			\mathcal{L} = &\mathcal{L}_{r} + \mathcal{L}_{e} + \lambda_{1}\left(\mathcal{L}_{re}^{s}+\mathcal{L}_{re}^{t} \right ) + 
			\lambda_{2} \mathcal{L}_{sty} + \\
			&\lambda_{3} \left ( \mathcal{L}_{r}^{adv} + \mathcal{L}_{e}^{adv} \right ),
		\end{aligned}
		\label{total_loss}
	\end{equation}
	where $\lambda_{1}$, $\lambda_{2}$ and $\lambda_{3}$ are different weighting coefficients. For this study, we empirically set $\lambda_{1}$ to $0.1$~\cite{article32}, $\lambda_{2}$ to $0.001$~\cite{article68} and $\lambda_{3}$ to $0.05$~\cite{article69}.
	
	\subsection{Implementation Details}
	Due to the obvious physiological characteristics of OD~\cite{article11}, it is easy to locate OD on a fundus image. We focused only on the segmentation of OD and OC in the cropped ROI. Given a fundus image, we first cropped a ROI of size $512\times512$ around OD and resized it to $256\times256$ for computational efficiency~\cite{article13}. Considering the limited number of training samples, we resorted common data augmentation strategies, including random scaling, random rotation, random flip, elastic transformation, adding salt-pepper noise, random erasing, and brightness adjustment~\cite{article11}, to diversify the training set. Our RDR-Net and all competing methods used the same set of augmentation strategies without the help of any extra post-processing. We implemented RDR-Net using the PyTorch framework on a workstation with one NVIDIA 1080Ti GPU.
	
	On the training phase, we optimized the segmentation network and discriminators in a two-step iterative way. We adopted the Adam optimizer for the network and the SGD optimizer for the discriminators. We set the batch size to 8, the learning rate of the network to $0.001$ with a decay of $0.1$ for every 100 epochs during the 200 epochs, and the learning rate of discriminators to $2.5e^{-5}$ without a decay.

	\subsection{Evaluation Metrics}
	For this study, the performance of OD segmentation and OC segmentation was measured separately by the Dice coefficient (Dice), mean Intersection over Union (mIoU), and pixel-wise Accuracy (Acc). A higher value of Dice, mIoU, or Acc means better performance.
	The performance of interactive segmentation between OD and OC was evaluated by the mean absolute error of CDR, which is commonly used in clinic practice to measure the optic abnormality.
	Let $\lbrace \hat{d}_c, \hat{d}_d\rbrace$ and $\lbrace d_c, d_d\rbrace$ denote the vertical cup diameter and vertical disc diameter obtained on a segmentation result and the corresponding ground truth, respectively. The CDR, denoted by $\delta$, is calculated as 
	\begin{equation}
		\small
		\begin{aligned}
			\delta = \left | \frac{\hat{d}_c}{\hat{d}_d}-\frac{d_c}{d_d} \right |.
		\end{aligned}
	\end{equation}
	A lower value of $\delta$ means better segmentation performance.

	\section{Experiments}
	\label{sec:Experiments}
	\subsection{Dataset}
	
	\begin{table*}[!htb]
		\caption{Statistics of four fundus image datasets used for this study.}
		\setlength{\tabcolsep}{3pt}
		\centering
		\begin{tabular}{m{80pt}<{\centering} m{95pt}<{\centering} m{95pt}<{\centering} m{95pt}<{\centering} m{95pt}<{\centering}}
			\hline
			Datasets&Drishti-GS&RIM-ONE-r3&REFUGE (Train)&ORIGA\\
			\hline
			Resolution&2047×1760&1072×1424&2124×2056&3072×2048\\
			
			Camera device&unknown&Canon EOS 5D&Zeiss Viscucam 50&unknown\\
			
			Number of images&50 Train + 51 Test&99 Train + 60 Test&400 Train + 0 Test&500 Train + 150 Test\\
			
			Year of Release &2014&2015&2018&2017\\
			\hline
		\end{tabular}
		\label{tab1}
	\end{table*}
	
	Four public fundus image datasets were used for this study, including the Drishti-GS dataset~\cite{article43}, RIM-ONE-r3 dataset~\cite{article44}, REFUGE dataset (only training set)~\cite{article45} and ORIGA dataset~\cite{article46}. The statistics of these datasets were listed in Table~\ref{tab1}. 
	To ensure a fair comparison, we followed the experimental settings used in~\cite{article11,article12,article13}.
	Due to the lack of an official split of the ORIGA dataset, we sorted the images by their file names from small to large and chose the first 500 images for training and the rest for test~\cite{article13}.
	
	\subsection{Results}
	First, we used the training set of REFUGE as the source domain and used Drishti-GS and RIM-ONE-r3 as the target domain, respectively. 
	We compared the proposed RDR-Net with five domain adaptation methods and two baseline settings: 'No Adapt' (\emph{i.e.}, training only on the source domain and test on the target domain), 'Upper bound' (\emph{i.e.}, training and test on the same target domain). 
	The UDA method proposed by Hoffman \emph{et al.}~\cite{article52} uses the global domain alignment. The method proposed by Javanmardi \emph{et al.}~\cite{article51} adopts adversarial learning to alleviate the domain shift issue. $\rho$OSAL~\cite{article11}, BEAL~\cite{article12}, and ISFA~\cite{article13} are three state-of-the-arts UDA methods for joint OD and OC segmentation on fundus images. All of them employ adversarial learning to align features, and ISFA also uses an additional CycleGAN to transfer images from the source domain to the target domain.
	The segmentation performance of these methods was reported in Table~\ref{tab2}. Among them, the performance of five completing methods were adopted from~\cite{article13}. The best results were highlighted in bold. It shows that ISFA and our RDR-Net, which jointly use image reconstruction and adversarial learning, substantially outperform BEAL and $\rho$OSAL, which only use adversarial learning. This observation confirms that using image reconstruction to shorten the distribution distance between the images from different domains can improve the segmentation performance.
	It also shows that, due to the larger domain discrepancy between the REFUGE training set and RIM-ONE-r3, all the methods perform worse on RIM-ONE-r3 than on Drishti-GS.
	Nevertheless, our RDR-Net achieves similar results to ISFA on Drishti-GS, but gains $1.0\%$ and $1.8\%$ $Dice$ improvement for disc and cup segmentation on RIM-ONE-r3. These results suggest that our RDR-Net is able to gain advantages from refining the low-level features, becoming particularly effective in handling domain discrepancy.
	
	\begin{table}[!tb]
		\caption{Performance of two baseline settings, five UDA methods and our RDR-Net in OD/OC segmentation, when using REFUGE training set as source domain and using Drishti-GS and RIM-ONE-r3 as target domain, respectively.}
		\setlength{\tabcolsep}{3pt}
		\centering
		\scriptsize
		\begin{tabular}{m{15pt}<{\centering}m{10pt}<{\centering}m{15pt}<{\centering}m{15pt}<{\centering}m{15pt}<{\centering}m{15pt}<{\centering}m{15pt}<{\centering}m{15pt}}
			\toprule
			\multicolumn{2}{c|}{Method} & \multicolumn{3}{c|}{Drishti-GS} & \multicolumn{3}{c}{RIM-ONE-r3}  
			\\ \cline{3-8} 
			\multicolumn{2}{c|}{} & \multicolumn{1}{c|}{$Dice_{OD}$} & \multicolumn{1}{c|}{$Dice_{OC}$} & \multicolumn{1}{c|}{$\delta$} & \multicolumn{1}{c|}{$Dice_{OD}$} & \multicolumn{1}{c|}{$Dice_{OC}$} & \multicolumn{1}{c}{$\delta$} 
			\\ \hline
			\multicolumn{2}{c|}{No Adapt} & \multicolumn{1}{c|}{$0.952$} & \multicolumn{1}{c|}{$0.842$} & \multicolumn{1}{c|}{$0.105$} & \multicolumn{1}{c|}{$0.855$} & \multicolumn{1}{c|}{$0.769$} & $0.086$ 
			\\
			\multicolumn{2}{c|}{Upper bound} & \multicolumn{1}{c|}{$0.977$} & \multicolumn{1}{c|}{$0.910$} & \multicolumn{1}{c|}{$0.039$} & \multicolumn{1}{c|}{$0.969$} & \multicolumn{1}{c|}{$0.877$} & $0.041$ 
			\\
			\hline
			\multicolumn{2}{c|}{Hoffman~\textsl{et al.}~\cite{article52}} & \multicolumn{1}{c|}{$0.959$} & \multicolumn{1}{c|}{$0.851$} & \multicolumn{1}{c|}{$0.093$} & \multicolumn{1}{c|}{$0.852$} & \multicolumn{1}{c|}{$0.755$} & $0.082$ \\ 
			\multicolumn{2}{c|}{\fontsize{6.5pt}{\baselineskip}\selectfont{Javanmardi~\textsl{et al.}~\cite{article51}}} & \multicolumn{1}{c|}{$0.961$} & \multicolumn{1}{c|}{$0.849$} & \multicolumn{1}{c|}{$0.091$} & \multicolumn{1}{c|}{$0.853$} & \multicolumn{1}{c|}{$0.779$} & $0.085$ \\ 
			\multicolumn{2}{c|}{$\rho$OSAL~\cite{article11}} & \multicolumn{1}{c|}{$0.965$} & \multicolumn{1}{c|}{$0.858$} & \multicolumn{1}{c|}{$0.082$} & \multicolumn{1}{c|}{$0.865$} & \multicolumn{1}{c|}{$0.787$} & $0.081$ \\ 
			\multicolumn{2}{c|}{BEAL~\cite{article12}} & \multicolumn{1}{c|}{$0.961$} & \multicolumn{1}{c|}{$0.862$} & \multicolumn{1}{c|}{-} & \multicolumn{1}{c|}{$0.898$} & \multicolumn{1}{c|}{$0.810$} & - \\ 
			\multicolumn{2}{c|}{ISFA~\cite{article13}} & \multicolumn{1}{c|}{$0.966$} & \multicolumn{1}{c|}{$0.892$} & \multicolumn{1}{c|}{-} & \multicolumn{1}{c|}{$0.908$} & \multicolumn{1}{c|}{$0.822$} & - \\ 
			\multicolumn{2}{c|}{Ours} & \multicolumn{1}{c|}{$\mathbf{0.971}$} & \multicolumn{1}{c|}{$\mathbf{0.893}$} & \multicolumn{1}{c|}{$\mathbf{0.062}$} & \multicolumn{1}{c|}{$\mathbf{0.918}$} & \multicolumn{1}{c|}{$\mathbf{0.840}$} & $\mathbf{0.059}$ \\ 
			\bottomrule
		\end{tabular}
		\label{tab2}
	\end{table}
	
	\begin{table*}[!htb]
		\caption{Performance of three UDA methods and our RDR-Net in OD/OC segmentation, when using either Drishti-GS or RIM-ONE-r3 as source domain and using others as target domains. The p-values were calculated based on the mean Dice of RDR-Net and each completing model.}
		\setlength{\tabcolsep}{3pt}
		\centering
		\scriptsize
		\begin{tabular}{m{40pt}<{\centering}|m{40pt}<{\centering}|m{45pt}<{\centering}|m{40pt}<{\centering}|m{40pt}<{\centering}|m{40pt}<{\centering}|m{40pt}<{\centering}|m{40pt}<{\centering}|m{40pt}<{\centering}|m{35pt}<{\centering}|m{35pt}<{\centering}}
			\toprule
			Source & Target & Method & $Dice_{OD}$ & $Dice_{OC}$ & $mIoU_{OD}$ & $mIoU_{OC}$ & $Acc_{OD}$ & $Acc_{OC}$ & $\delta$ & p-value \\
			\hline
			\multicolumn{1}{c|}{\multirow{8}{*}{Drishti-GS}} & \multirow{4}{*}{RIM-ONE-r3} & $\rho$OSAL~\cite{article11} & $0.9153$ & $0.8159$ & $0.8957$ & $0.8383$ & $0.9564$ & $0.9790$ & $0.0692$ & $0.004$ 
			\\
			\multicolumn{1}{c|}{ } & & BEAL~\cite{article12} & $0.9208$ & $0.8001$ & $0.9003$ & $0.8345$ & $0.9591$ & $0.9788$ & $0.0753$ & $<0.001$ 
			\\
			\multicolumn{1}{c|}{ } & & ISFA~\cite{article13} & $\mathbf{0.9252}$ & $0.8220$ & $\mathbf{0.9059}$ & $0.8471$ & $\mathbf{0.9616}$ & $0.9804$ & $\backslash$ & $\backslash$ 
			\\
			\multicolumn{1}{c|}{ } & & Ours & $0.9215$ & $\mathbf{0.8282}$ & $0.9028$ & $\mathbf{0.8512}$ & $0.9607$ & $\mathbf{0.9820}$ & $\mathbf{0.0566}$ & $\backslash$ 
			\\ \cline{2-11}
			\multicolumn{1}{c|}{ } & \multirow{4}{*}{ORIGA} & $\rho$OSAL~\cite{article11} & $0.9445$ & $0.8498$ & $0.9186$ & $0.8508$ & $0.9540$ & $0.9568$ & $0.0843$ & $<0.001$ 
			\\
			\multicolumn{1}{c|}{ } & & BEAL~\cite{article12} & $0.9490$ & $0.8524$ & $0.9204$ & $0.8525$ & $0.9564$ & $0.9580$ & $0.0810$ & $<0.001$ 
			\\
			\multicolumn{1}{c|}{ } & & ISFA~\cite{article13} & $0.9587$ & $0.8728$ & $0.9339$ & $0.8676$ & $0.9674$ & $0.9616$ & $\backslash$ & $\backslash$ \\	 
			\multicolumn{1}{c|}{ } & & Ours & $\mathbf{0.9684}$ & $\mathbf{0.8834}$ & $\mathbf{0.9494}$ & $\mathbf{0.8792}$ & $\mathbf{0.9755}$ & $\mathbf{0.9656}$ & $\mathbf{0.0775}$ & $\backslash$ \\	 	
			\hline
			
			\multicolumn{1}{c|}{\multirow{8}{*}{RIM-ONE-r3}} & \multirow{4}{*}{Drishti-GS} & $\rho$OSAL~\cite{article11} & $0.9096$ & $0.8303$ & $0.8614$ & $0.8325$ & $0.9294$ & $0.9497$ & $0.0761$ & $0.001$ 
			\\
			\multicolumn{1}{c|}{ } & & BEAL~\cite{article12} & $0.9293$ & $0.8054$ & $0.8927$ & $0.8116$ & $0.9478$ & $0.9435$ & $0.0878$ & $0.001$ 
			\\
			\multicolumn{1}{c|}{ } & & ISFA~\cite{article13} & $0.9324$ & $0.8385$ & $0.8961$ & $0.8413$ & $0.9492$ & $0.9549$ & $\backslash$ & $\backslash$\\
			\multicolumn{1}{c|}{ } & & Ours & $\mathbf{0.9434}$ & $\mathbf{0.8430}$ & $\mathbf{0.9124}$ & $\mathbf{0.8455}$ & $\mathbf{0.9575}$ & $\mathbf{0.9552}$ & $\mathbf{0.0638}$ & $\backslash$\\ \cline{2-11}
			
			\multicolumn{1}{c|}{ } & \multirow{4}{*}{ORIGA} & $\rho$OSAL~\cite{article11} & $0.9072$ & $0.8068$ & $0.8740$ & $0.8135$ & $0.9361$ & $0.9489$ & $0.0885$ & $<0.001$ 
			\\
			\multicolumn{1}{c|}{ } & & BEAL~\cite{article12} & $0.9369$ & $0.8378$ & $0.9040$ & $0.8398$ & $0.9528$ & $0.9546$ & $0.0938$ & $0.005$ 
			\\
			\multicolumn{1}{c|}{ } & & ISFA~\cite{article13} & $0.9481$ & $0.8449$ & $0.9192$ & $0.8439$ & $0.9604$ & $0.9558$ & $\backslash$ & $\backslash$ \\
			\multicolumn{1}{c|}{ } & & Ours & $\mathbf{0.9521}$ & $\mathbf{0.8463}$ & $\mathbf{0.9256}$ & $\mathbf{0.8516}$ & $\mathbf{0.9635}$ & $\mathbf{0.9608}$ & $\mathbf{0.0806}$ & $\backslash$ \\
			\bottomrule
		\end{tabular}			
		\label{tab6}
	\end{table*}
	
	Second, to evaluate the performance of our model trained on a small source dataset, we used either Drishti-GS or RIM-ONE-r3 as the source domain and used the other one and ORIGA as the target domain, respectively.
	The performance of our RDR-Net and three state-of-the-art UDA methods (\emph{i.e.}, $\rho$OSAL, BEAL, and ISFA) was given in Table~\ref{tab6}. The best results were highlighted in bold. Note that the performance of ISFA is directly adopted from~\cite{article13}, where the experimental settings are the same to those used for this study. The results of $\rho$OSAL and BEAL were obtained by reproducing the codes.
	
\begin{figure*}[!tb]
		\centering
		\includegraphics[width=\textwidth]{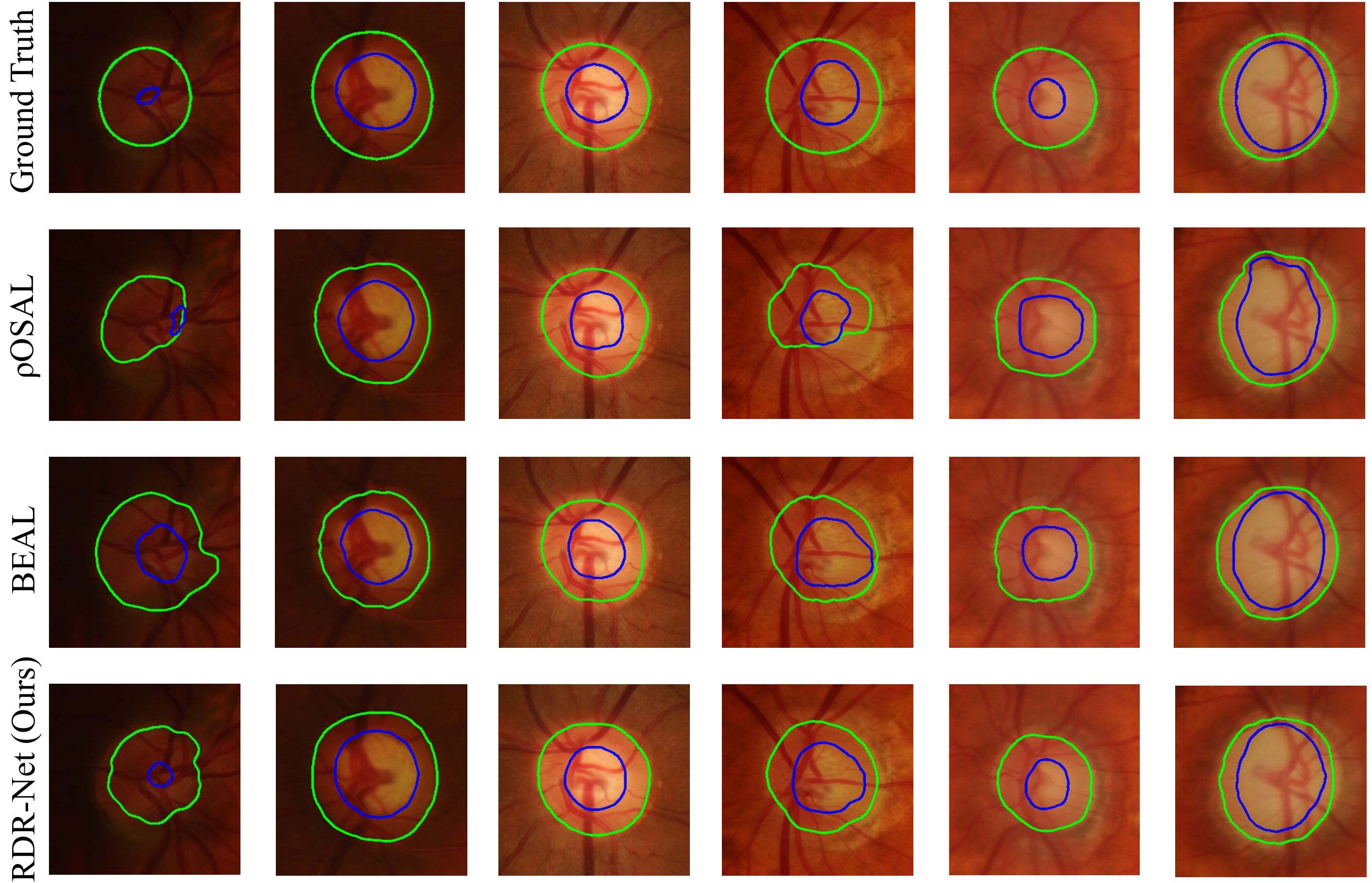}
		\caption{Visualization of OD (green) and OC (blue) segmentation results obtained on RIM-ONE-r3 test set when using Drishti-GS as source domain. From top to bottom: Ground truth of OD and OC (1st row), and segmentation results of $\rho$OSAL (2nd row), BEAL (3rd row) and our RDR-Net (4th row).}
		\label{fig9}
	\end{figure*}

	It reveals that our RDR-Net is substantially superior to $\rho$OSAL and BEAL in most cases, especially when the target dataset is large (\emph{e.g.}, ORIGA). It can be attributed to the fact that the reconstruction alignment can provide more self-supervised guidance from a large number of images~\cite{article67}.
	Comparing to ISFA, RDR-Net achieves competitive performance in OD segmentation and superior performance in OC segmentation. Since both ISFA and RDR-Net jointly use image reconstruction and adversarial learning, we believe that the performance gain of RDR-Net stems from the dynamic convolution-based low-level feature refinement.
	We also conducted statistical tests on the mean value of $Dice_{OD}$ and $Dice_{OC}$, and reported the $p$-values. All $p$-values are smaller than 0.05, suggesting that the performance gain of our RDR-Net over each competing method is statistically significant.
	Moreover, Fig.~\ref{fig9} gives the visualization of some results on the RIM-ONE-r3 dataset while using Drishti-GS as the source domain. It shows that our RDR-Net can produce more accurate segmentation results than other methods, particularly on difficult examples (see the 1st and 4th column).

	\begin{table*}[!hbt]
		\caption{Ablation study on different components. The p-values were calculated based on the mean Dice of RDR-Net (\emph{i.e.}, the baseline with all components) and each variant model.}
		\setlength{\tabcolsep}{3pt}
		\centering
		\scriptsize
		\begin{tabular}{m{30pt}<{\centering}|m{30pt}<{\centering}|m{30pt}<{\centering}|m{25pt}<{\centering}|m{30pt}<{\centering}|m{25pt}<{\centering}|m{40pt}<{\centering}|m{40pt}<{\centering}|m{40pt}<{\centering}|m{40pt}<{\centering}|m{40pt}<{\centering}|m{40pt}<{\centering}|m{35pt}<{\centering}}
			\toprule
			\multicolumn{2}{c|}{Target Domain} & Baseline & RA & LFR & \multicolumn{1}{c|}{PMA} & $Dice_{OD}$ & $Dice_{OC}$ & $mIoU_{OD}$ & $mIoU_{OC}$ & $Acc_{OD}$ & $Acc_{OC}$ & p-value \\
			\hline
			\multicolumn{2}{c|}{ } & $\checkmark$ & & & & $0.8549$ & $0.7690$ & $0.8551$ & $0.8148$ & $0.9293$ & $0.9684$ & $<0.001$ 
			\\
			\multicolumn{2}{c|}{ } & $\checkmark$ & $\checkmark$ & & & $0.9033$ & $0.8289$ & $0.8786$ & $0.8511$ & $0.9490$ & $0.9783$ & $<0.001$ 
			\\
			\multicolumn{2}{c|}{\multirow{4}*{RIM-ONE-r3}} & $\checkmark$ & & $\checkmark$ & & $0.8990$ & $0.8262$ & $0.8731$ & $0.8509$ & $0.9460$ & $0.9786$ & $<0.001$ 
			\\
			\multicolumn{2}{c|}{ } & $\checkmark$ & & & $\checkmark$ & $0.8984$ & $0.8150$ & $0.8723$ & $0.8435$ & $0.9455$ & $0.9765$ & $<0.001$ 
			\\
			
			\multicolumn{2}{c|}{ } & $\checkmark$ &$\checkmark$ & & $\checkmark$ & $0.9064$ & $0.8315$ & $0.8816$ & $0.8518$ & $0.9499$ & $0.9787$ & $<0.001$ \\
			\multicolumn{2}{c|}{ } & $\checkmark$ & &$\checkmark$ & $\checkmark$ & $0.9067$ & $0.8325$ & $0.8824$ & $0.8521$ & $0.9506$ & $0.9789$ & $0.001$ 
			\\
			\multicolumn{2}{c|}{ } & $\checkmark$ & $\checkmark$ & $\checkmark$ & & $0.9110$ & $0.8346$ & $0.8918$ & $0.8536$ & $0.9559$ & $0.9792$ & $0.003$ 
			\\
			\multicolumn{2}{c|}{ } & $\checkmark$ & $\checkmark$ & $\checkmark$ & $\checkmark$ & $\mathbf{0.9179}$ & $\mathbf{0.8402}$ & $\mathbf{0.9098}$ & $\mathbf{0.8587}$ & $\mathbf{0.9641}$ & $\mathbf{0.9810}$ & $\backslash$ \\
			\hline
			\multicolumn{2}{c|}{ } & $\checkmark$ & & & & $0.9521$ & $0.8423$ & $0.9204$ & $0.8416$ & $0.9602$ & $0.9442$ & $<0.001$ 
			\\
			\multicolumn{2}{c|}{ } & $\checkmark$ & $\checkmark$ & & & $0.9615$ & $0.8807$ & $0.9414$ & $0.8771$ & $0.9693$ & $0.9612$ & $<0.001$ 
			\\
			\multicolumn{2}{c|}{\multirow{4}*{Drishti-GS}} & $\checkmark$ & & $\checkmark$ & & $0.9596$ & $0.8790$ & $0.9401$ & $0.8715$ & $0.9679$ & $0.9603$ & $<0.001$ 
			\\
			\multicolumn{2}{c|}{ } & $\checkmark$ & & & $\checkmark$ & $0.9604$ & $0.8683$ & $0.9366$ & $0.8622$ & $0.9675$ & $0.9573$ & $<0.001$ 
			\\
			\multicolumn{2}{c|}{ } & $\checkmark$ &$\checkmark$ & & $\checkmark$ & $0.9628$ & $0.8821$ & $0.9418$ & $0.8803$ & $0.9704$ & $0.9614$ & $<0.001$ \\
			\multicolumn{2}{c|}{ } & $\checkmark$ & &$\checkmark$ & $\checkmark$ & $0.9635$ & $0.8847$ & $0.9429$ & $0.8828$ & $0.9713$ & $0.9619$ & $<0.001$ \\
			\multicolumn{2}{c|}{ } & $\checkmark$ & $\checkmark$ & $\checkmark$ & & $0.9651$ & $0.8884$ & $0.9452$ & $0.8878$ & $0.9723$ & $0.9630$ & $0.004$ 
			\\
			\multicolumn{2}{c|}{ } & $\checkmark$ & $\checkmark$ & $\checkmark$ & $\checkmark$ & $\mathbf{0.9712}$ & $\mathbf{0.8934}$ & $\mathbf{0.9520}$ & $\mathbf{0.8930}$ & $\mathbf{0.9765}$ & $\mathbf{0.9694}$ & $\backslash$ \\
			\bottomrule
		\end{tabular}			
		\label{tab3}
	\end{table*}
	
	\begin{figure*}[!tb]
		\centering
		\includegraphics[width=\textwidth]{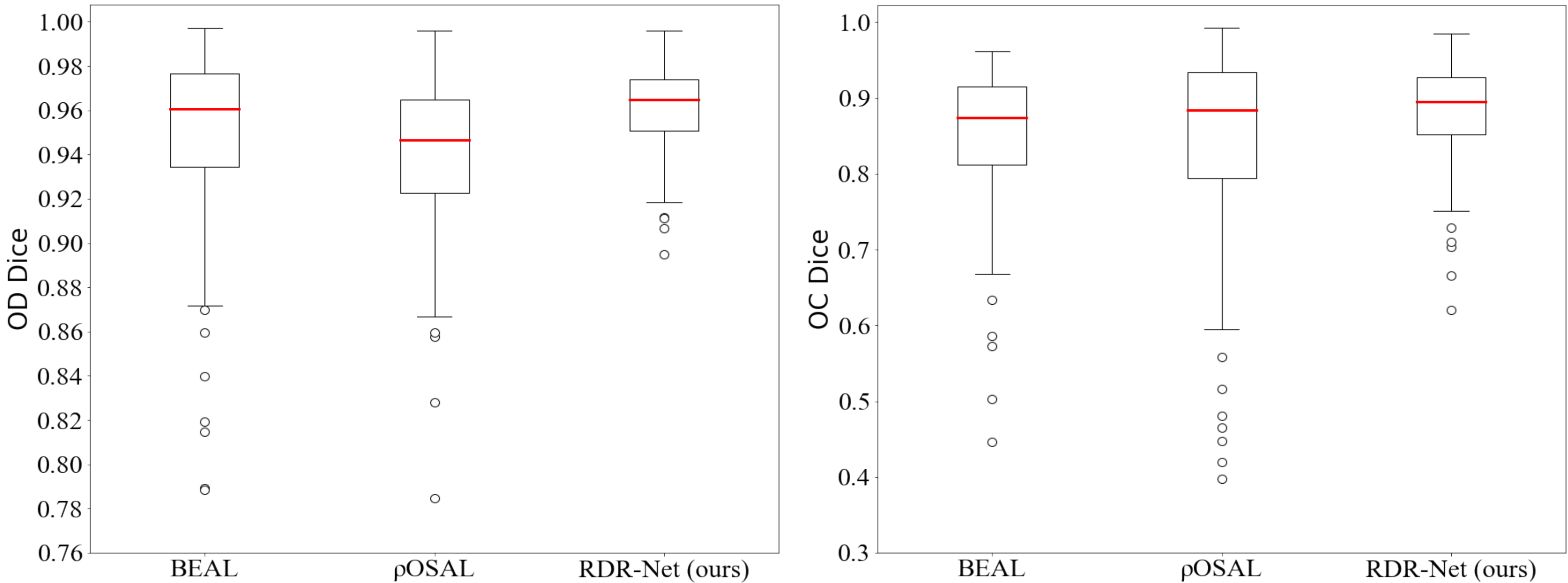}
		\caption{Boxplot of Dice scores of BEAL, $\rho$OSAL, and our RDR-Net when using REFUGE training set as source domain and ORIGA as target domain.}
		\label{fig-boxplot}
	\end{figure*}
	
	\subsection{Ablation Study}
	We designed three key modules (\emph{i.e.}, RA, LFR, and PMA) to enable our RDR-Net to address the domain gap issue for better OD/OC segmentation. To evaluate the effectiveness of each module, we performed ablation studies, in which REFUGE was used as the source domain and Drishti-GS and RIM-ONE-r3 were used as two target domains. We compared the performance of the baseline (w/o UDA), baseline+RA, baseline+LFR, baseline+PMA, baseline+RA+PMA, baseline+LFR+PMA, baseline+RA+LFR, and baseline+RA+LFR+PMA (\emph{i.e.}, our RDR-Net). Note that ``baseline+LFR" only utilizes the high-level features $F_h$ compressed by GAP to refine the low-level features $F_l$ without using the distribution information produced by RA.
	The results were reported in Table~\ref{tab3}. 
	It shows that adding each of RA, LFR, and PMA to the baseline can improve the performance of OC/OD segmentation on both target domains. Particularly, the performance gain caused by RA or LFR is much larger than that caused by PMA, suggesting that learning feature representations on the target domain via reconstruction-based self-supervised learning and refining $F_l$ with the knowledge from $F_h$ are more effective in UDA than using adversarial learning.
	Meanwhile, it also reveals that jointly using any two modules further improves the segmentation performance and using the combination of all three modules leads to the best performance.
	Moreover, although LFR alone is inferior to RA, the combination of LFR and PMA is superior to the combination of RA and PMA. It can be attributed that the high-level features benefit from PMA a lot and thus provide better guidance for the LFR module.
	To verify the significance, we also conducted the Wilcoxon rank-sum test on the mean value of OC Dice and OD Dice obtained by our RDR-Net and each of other combinations. It shows that the $p$-values are all less than 0.05. 
	The results of our ablation studies indicate that each of the RA, LFR, and PMA modules is effective and the performance gain caused by each module is statistically significant.
	
	\subsection{Performance Stability}
	We chose the REFUGE training set as the source domain and ORIGA as the target domain and used this setting as a case study to evaluate the stability of our RDR-Net against BEAL and $\rho$OSAL. The boxplots of the Dice scores obtained by three models were displayed in Fig~\ref{fig-boxplot}. It shows that our RDR-Net has the largest median and smallest quartile, indicating better performance stability of RDR-Net. Moreover, the outliers of RDR-Net have higher Dice scores than those of other two networks, suggesting the effectiveness of our RDR-Net on hard samples.

	\begin{table}[!tb]
		\caption{Comparison on refining low- or high-level features.}
		\setlength{\tabcolsep}{3pt}
		\centering
		\scriptsize
		\begin{tabular}{m{30pt}<{\centering}m{25pt}<{\centering}m{30pt}<{\centering}m{30pt}<{\centering}m{25pt}<{\centering}m{25pt}<{\centering}m{25pt}<{\centering}m{25pt}<{\centering}}
			\toprule
			\multicolumn{2}{c|}{LFR} & 
			\multicolumn{2}{c|}{Concat} & \multicolumn{2}{c|}{Drishti-GS} & \multicolumn{2}{c}{RIM-ONE-r3} \\ \cline{5-8} 
			\multicolumn{2}{c|}{} &
			\multicolumn{2}{c|}{} &
			\multicolumn{1}{c|}{$Dice_{OD}$} & \multicolumn{1}{c|}{$Dice_{OC}$} & \multicolumn{1}{c|}{$Dice_{OD}$} & \multicolumn{1}{c}{$Dice_{OC}$} \\ \hline
			\multicolumn{2}{c|}{Low} &
			\multicolumn{2}{c|}{} &
			\multicolumn{1}{c|}{$0.9665$} & \multicolumn{1}{c|}{$0.8822$} & \multicolumn{1}{c|}{$0.8942$} & \multicolumn{1}{c}{$0.8220$} \\ 
			\multicolumn{2}{c|}{Low} &
			\multicolumn{2}{c|}{\checkmark} &
			\multicolumn{1}{c|}{$\mathbf{0.9712}$} & \multicolumn{1}{c|}{$\mathbf{0.8934}$} & \multicolumn{1}{c|}{$\mathbf{0.9179}$} & \multicolumn{1}{c}{$\mathbf{0.8402}$} \\ 
			\multicolumn{2}{c|}{High} &
			\multicolumn{2}{c|}{} &
			\multicolumn{1}{c|}{$0.9639$} & \multicolumn{1}{c|}{$0.8708$} & \multicolumn{1}{c|}{$0.8910$} & \multicolumn{1}{c}{$0.8178$} \\ 
			\multicolumn{2}{c|}{High} &
			\multicolumn{2}{c|}{\checkmark} &
			\multicolumn{1}{c|}{$0.9673$} & \multicolumn{1}{c|}{$0.8871$} & \multicolumn{1}{c|}{$0.8955$} & \multicolumn{1}{c}{$0.8268$} \\ 
			\bottomrule
		\end{tabular}
		\label{tab7}
	\end{table}
	
	\begin{figure}[!htb]
		\centering
		\includegraphics[width=\columnwidth]{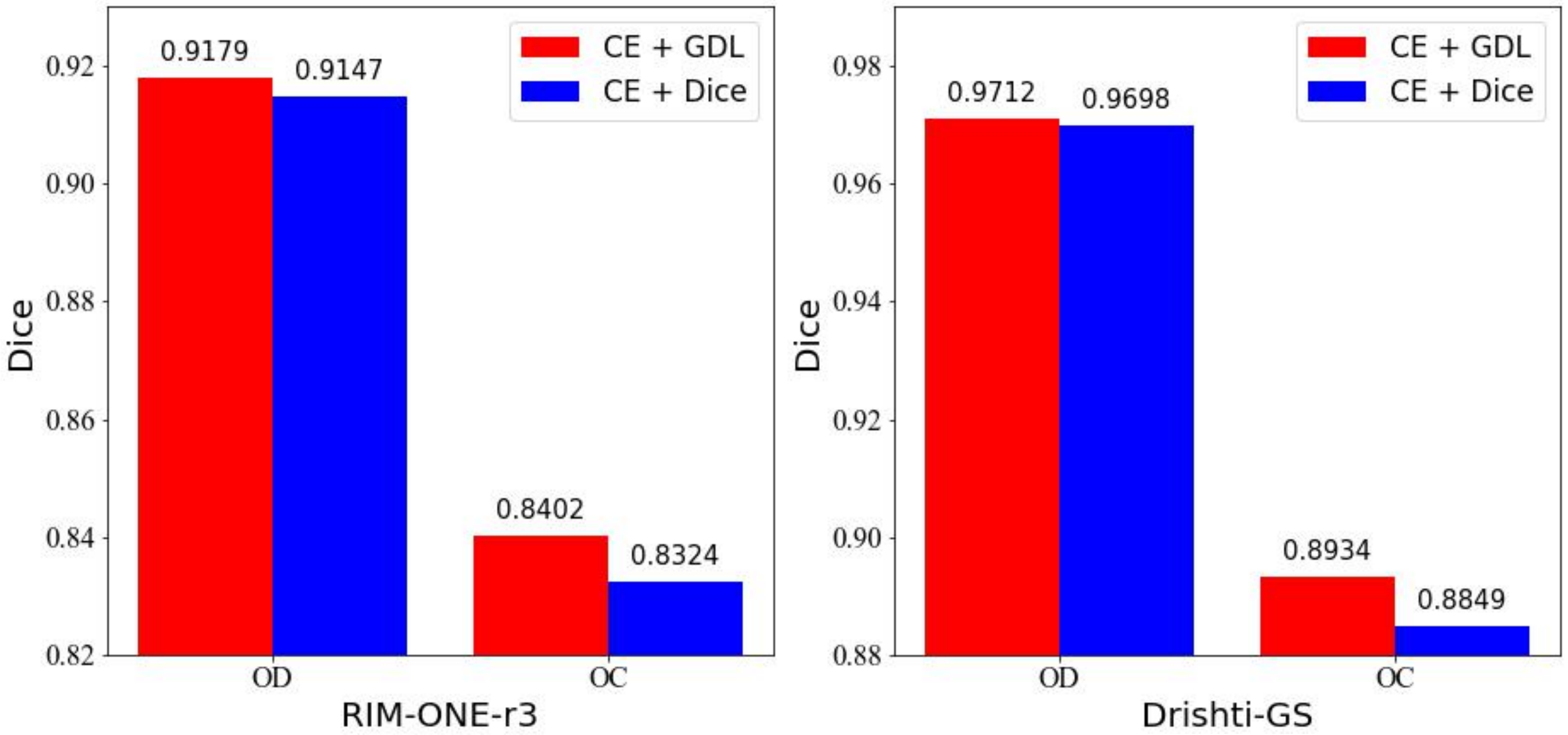}
		\caption{Comparison of different loss functions of segmentation.}
		\label{fig-loss}
	\end{figure}

	\begin{table}[!tb]
		\caption{Comparison on different values of weighting coefficient $\lambda_{1}$, $\lambda_{2}$ and $\lambda_{3}$.}
		\setlength{\tabcolsep}{3pt}
		\centering
		\scriptsize
		\begin{tabular}{c|cc|c|cc|c|cc}
			\toprule
			\multicolumn{1}{c|}{\multirow{3}*{$\lambda_{1}$}} & \multicolumn{2}{c|}{RIM-ONE-r3} & 
			\multicolumn{1}{c|}{\multirow{3}*{$\lambda_{2}$}} & \multicolumn{2}{c|}{RIM-ONE-r3} & 
			\multicolumn{1}{c|}{\multirow{3}*{$\lambda_{3}$}} & \multicolumn{2}{c}{RIM-ONE-r3} \\
			
			\multicolumn{1}{c|}{} & \multicolumn{2}{c|}{$\lambda_2$=0.001, $\lambda_3$=0.05} & \multicolumn{1}{c|}{} & \multicolumn{2}{c|}{$\lambda_1$=0.1, $\lambda_3$=0.05} & \multicolumn{1}{c|}{} & \multicolumn{2}{c}{$\lambda_1$=0.1, $\lambda_2$=0.001}
			\\ \cline{2-3} \cline{5-6} \cline{8-9}
			
			\multicolumn{1}{c|}{} & \multicolumn{1}{c|}{${OD}$} & \multicolumn{1}{c|}{${OC}$} & \multicolumn{1}{c|}{} & \multicolumn{1}{c|}{${OD}$} & \multicolumn{1}{c|}{${OC}$} & \multicolumn{1}{c|}{} & \multicolumn{1}{c|}{${OD}$} & \multicolumn{1}{c}{${OC}$} \\ \hline
			
			\multicolumn{1}{c|}{0} & \multicolumn{1}{c|}{$.9120$} & \multicolumn{1}{c|}{$.8307$} & \multicolumn{1}{c|}{0.00001} & \multicolumn{1}{c|}{$.9139$} & \multicolumn{1}{c|}{$.8354$} & \multicolumn{1}{c|}{0.001} & \multicolumn{1}{c|}{$.9104$} & \multicolumn{1}{c}{$.8369$}\\ 
			
			\multicolumn{1}{c|}{0.01} & \multicolumn{1}{c|}{$.9149$} & \multicolumn{1}{c|}{$.8360$} & \multicolumn{1}{c|}{0.0001} & \multicolumn{1}{c|}{$.9161$} & \multicolumn{1}{c|}{$.8373$} & \multicolumn{1}{c|}{0.01} & \multicolumn{1}{c|}{$.9131$} & \multicolumn{1}{c}{$\mathbf{.8413}$} \\ 
			
			\multicolumn{1}{c|}{0.1} & \multicolumn{1}{c|}{$\mathbf{.9179}$} & \multicolumn{1}{c|}{$\mathbf{.8402}$} & \multicolumn{1}{c|}{0.001} & \multicolumn{1}{c|}{$\mathbf{.9179}$} & \multicolumn{1}{c|}{$\mathbf{.8402}$} & \multicolumn{1}{c|}{0.05} & \multicolumn{1}{c|}{$\mathbf{.9179}$} & \multicolumn{1}{c}{${.8402}$} \\ 
			
			\multicolumn{1}{c|}{1.0} & \multicolumn{1}{c|}{$.9099$} & \multicolumn{1}{c|}{$.8251$} & \multicolumn{1}{c|}{0.01} & \multicolumn{1}{c|}{$.9124$} & \multicolumn{1}{c|}{$.8377$} & \multicolumn{1}{c|}{0.1} & \multicolumn{1}{c|}{$.9087$} & \multicolumn{1}{c}{$.8339$} \\ 
			
			\multicolumn{1}{c|}{10.0} & \multicolumn{1}{c|}{$.9068$} & \multicolumn{1}{c|}{$.8231$} & \multicolumn{1}{c|}{0.01} & \multicolumn{1}{c|}{$.9043$} & \multicolumn{1}{c|}{$.8306$} & \multicolumn{1}{c|}{1.0} & \multicolumn{1}{c|}{$.8967$} & \multicolumn{1}{c}{$.8242$} \\ 
			\bottomrule
			\multicolumn{9}{l}{\footnotesize{$^{\rm 1}$The values are Dice scores, and each ``0" in ``0." is omitted.}}
		\end{tabular}
		\label{tab5}
	\end{table}
	
	\section{Discussion}
	\label{sec:Discussion}
	
	\subsection{Low- or High-level Feature Refinement}
	In the LFR module, we employed dynamic convolutions to refine low-level features $F_{l}$, aiming to suppress the domain-variant information in them. Since the high-level features $F_{h}$ contain more semantic information which are domain-invariant, refining $F_{h}$ in the same way is not cost-effective. To validate this, we performed the first experiment again to test the performance of our RDR-Net with either the low-level feature refinement or high-level feature refinement. We also evaluated the contributions made by concatenating the low-level and high-level features for segmentation. The results were reported in Table~\ref{tab7}. It shows that, no matter using feature concatenation or not, the model with low-level feature refinement outperforms the one with high-level feature refinement. As expected, the highest Dice values were achieved when using both low-level feature refinement and low- and high-level feature concatenation. Our results suggest that the proposed feature refinement should be applied to low-level features.

	\subsection{GDL vs. Dice Loss}
	The proposed RDR-Net was designed to segment both OD and OC on fundus images, where OC is always located inside OD and is, of course, smaller than OD. To make use of this prior knowledge, we replaced the traditional Dice loss with its generalized version, \emph{i.e.}, GDL, which controls the contribution that each class makes to the loss by weighting classes by the inverse size of the expected region~\cite{article40}.
	We minimized the sum of cross-entropy loss and GDL to optimize the segmentation backbone.
	To validate the effectiveness of GDL, we attempted to use the Dice loss and re-performed the first experiment. The results were displayed in Fig.~\ref{fig-loss}. 
	It shows that using the combination of cross-entropy loss and GDL leads to better segmentation performance than using the combination of cross-entropy loss and Dice loss, particularly in OC segmentation. Our results suggesting that GDL is more suitable than its traditional counterpart for this imbalanced segmentation task.

	\subsection{Value of Weighting Coefficients}
	As shown in Eq.~(\ref{total_loss}), the weighting coefficients $\lambda_{1}$, $\lambda_{2}$ and $\lambda_{3}$ control the contributions made by the image reconstruction, style-consistency constraint, and adversarial learning, respectively. To investigate the settings of three weighting coefficients, we repeated the experiments on the REFUGE training set (source) and RIM-ONE-r3 (target) and reported the performance of our RDR-Net with different parameter settings in Table~\ref{tab5}. It reveals that RDR-Net achieves the most accurate segmentation of both OD and OC when setting $\lambda_{1}$ to 0.1, $\lambda_{2}$ to 0.001, and $\lambda_{3}$ to 0.05.
	
	\begin{table}[!tb]
		\caption{Number of parameters, GFLOPs, training time, and inference time (on one sample) of different models.}
		\centering
		\scriptsize
		\begin{tabular}{m{35pt}<{\centering}m{25pt}<{\centering}m{25pt}<{\centering}m{25pt}<{\centering}m{25pt}<{\centering}}
			\toprule
			\multicolumn{1}{c|}{\multirow{2}*{Models}} & \multicolumn{1}{c}{Parameters} & \multicolumn{1}{c}{\multirow{2}*{GFLOPs}} & \multicolumn{1}{c}{Training} & \multicolumn{1}{c}{Inference}
			\\ 
			\multicolumn{1}{c|}{} & \multicolumn{1}{c}{($\times 10^6$)} & & \multicolumn{1}{c}{Time (h)} & \multicolumn{1}{c}{Time (s)}
			\\ \cline{1-5}
			\multicolumn{1}{c|}{$\rho$OSAL} &
			\multicolumn{1}{c}{$5.81$} &
			\multicolumn{1}{c}{$6.64$} &
			\multicolumn{1}{c}{$1.25$} &
			\multicolumn{1}{c}{$0.021$}
			\\
			\multicolumn{1}{c|}{BEAL} &
			\multicolumn{1}{c}{$5.81$} &
			\multicolumn{1}{c}{$6.64$} &
			\multicolumn{1}{c}{$2.51$} &
			\multicolumn{1}{c}{$0.021$}
			\\
			\multicolumn{1}{c|}{Ours} &
			\multicolumn{1}{c}{$6.45$} &
			\multicolumn{1}{c}{$6.64$} &
			\multicolumn{1}{c}{$3.32$} &
			\multicolumn{1}{c}{$0.023$}
			\\ 
			\bottomrule
		\end{tabular}
		\label{tab:complexity}
	\end{table}
	
	\subsection{Complexity}
	The proposed RDR-Net contains a dual-path backbone and three modules. To evaluate its complexity, we chose the REFUGE training set as the source domain and Drishti-GS as the target domain and used this setting as a case study. Table~\ref{tab:complexity} gives the number of parameters, GFLOPs, training time, and inference time (on one 256 $\times$ 256 sample) of our RDR-Net, $\rho$OSAL, and BEAL. Note that, when calculating the number of parameters, GFLOPs, and inference time, only the first convolutional layer and fully connected layer in the VAE branch of RDR-Net are taken into account, since other layers in the VAE branch do not work in the inference phase. It shows that, although RDR-Net contains more parameters, it has similar GFLOPs, training time, and inference time with the other two models. The 
	efficiency mainly comes from that (1) the employed part of the VAE branch is operated only on the low-resolution feature maps, and (2) the dynamic convolutions are light-weighted. In summary, our RDR-Net is a little bit more complex than $\rho$OSAL and BEAL, but its training time (less than 4 hours) is still acceptable and its inference speed is very fast (less than 0.05 seconds per image).
	
	\begin{figure}[!tb]
		\centering
		\includegraphics[width=0.5\textwidth]{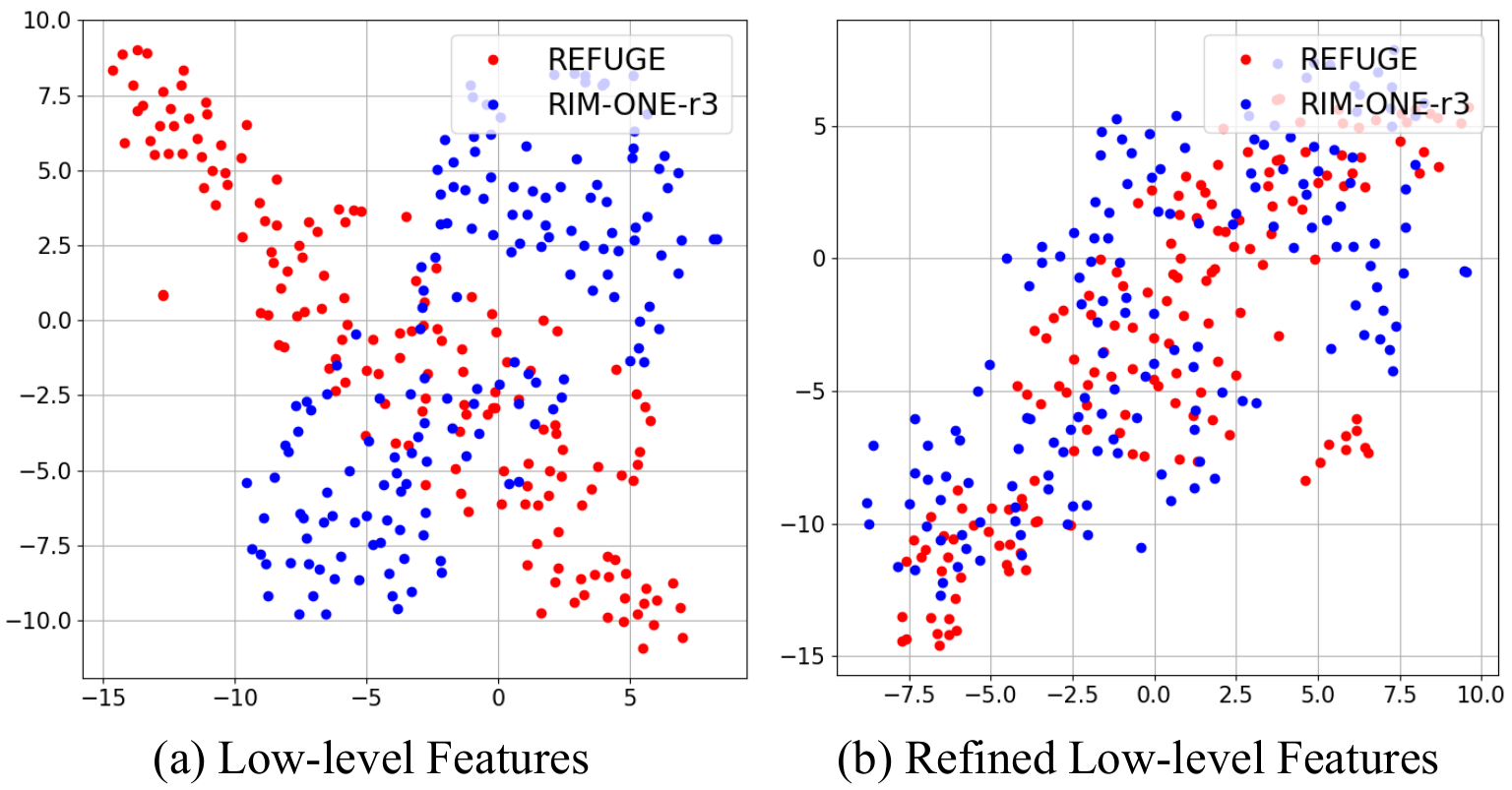}
		\caption{Visualization of low-level features before (left) and after (right) being refined by the LFR module using 2D t-SNE. The points in red and blue represent the samples from the source and target domains, respectively.}
		\label{fig-tsne}
	\end{figure}
	
	\subsection{Generalization Analysis of LFR}
	To verify the ability of our LFR module to filter the domain-variant information, we chose the model trained on the REFUGE training set (source domain) and RIM-ONE-r3 (target domain) as a case study. The low-level features before and after being refined by LFR were visualized using 2D t-SNE in Fig.~\ref{fig-tsne}. It reveals that the low-level features from two domains are largely separated from each other before being refined (see Fig.~\ref{fig-tsne}(a)) and become indistinguishable after being refined (see Fig.~\ref{fig-tsne}(b)). The visualization suggests that our LFR module can filter domain-variant information and can be generalizable to the target domain.
	\section{Conclusion}
	\label{sec:Conclusion}
	In this paper, we propose a UDA model called RDR-Net for OD/OC segmentation on fundus images. It addresses the domain shift issue by jointly using three strategies, including the VAE-based reconstruction alignment, dynamic convolution-based low-level feature refinement, and prediction-map alignment.
	The results obtained on four fundus datasets suggest that the proposed RDR-Net outperforms several UDA models on this medical image segmentation task and each strategy we developed is effective in delivering performance gains.
	
	However, the proposed RDR-Net still has two limitations. 
	First, it is designed for single-source UDA tasks, \emph{i.e.}, it can only use the data from a single source domain for training. When trained on multi-source datasets, the performance of RDR-Net may drop due to its inability to handle the domain gap within training data. 
	Second, VAE is beneficial to training, but not used in the inference phase, except for providing data distributions to the LFR module. Indeed, VAE-based image reconstruction provides the way to narrow down the domain gap at the image level.
	In our future work, we plan to incorporate such VAE-based image alignment into the UDA framework, aiming to address the domain gap issue from the image level, feature level, and decision level simultaneously.


	
	\newpage
	
	\section*{Appendix}
	\setcounter{table}{0}
	
	\subsection*{Analysis of Style-Consistency Loss}
	Recall that the style-consistency loss $\mathcal{L}_{sty}$ and reconstruction loss $\mathcal{L}_{re}$ are adversarial, we set a small weight for the former to limit its impact on the latter. To verify the effectiveness of $\mathcal{L}_{sty}$, we compared the performance of ``baseline+RA'' with and without $\mathcal{L}_{sty}$ in Table~\ref{A1}. It reveals that the RA module with $\mathcal{L}_{sty}$ outperforms the one without $\mathcal{L}_{sty}$, indicating the effectiveness of $\mathcal{L}_{sty}$.
	
	\begin{table}[!htb]
		\renewcommand{\thetable}{A\arabic{table}}
		\caption{Ablation study on the style-consistency loss, when using REFUGE training set as source domain and using Drishti-GS and RIM-ONE-r3 as target domains, respectively.}
		\setlength{\tabcolsep}{3pt}
		\centering
		\scriptsize
		\begin{tabular}{m{15pt}<{\centering}m{10pt}<{\centering}m{15pt}<{\centering}m{15pt}<{\centering}m{15pt}<{\centering}m{15pt}<{\centering}}
			\toprule
			\multicolumn{2}{c|}{Method} & \multicolumn{2}{c|}{Drishti-GS} & \multicolumn{2}{c}{RIM-ONE-r3} \\ \cline{3-6} 
			\multicolumn{2}{c|}{} & \multicolumn{1}{c|}{$Dice_{OD}$} & \multicolumn{1}{c|}{$Dice_{OC}$}  & \multicolumn{1}{c|}{$Dice_{OD}$} & \multicolumn{1}{c}{$Dice_{OC}$} \\ \hline
			\multicolumn{2}{c|}{baseline+RA (w/o $\mathcal{L}_{sty}$)} & \multicolumn{1}{c|}{$0.9521$} & \multicolumn{1}{c|}{$0.8718$}  & \multicolumn{1}{c|}{$0.8939$} & \multicolumn{1}{c}{$0.8233$} \\
			\multicolumn{2}{c|}{baseline+RA (w $\mathcal{L}_{sty}$)} & \multicolumn{1}{c|}{$\mathbf{0.9615}$} & \multicolumn{1}{c|}{$\mathbf{0.8807}$} & \multicolumn{1}{c|}{$\mathbf{0.9033}$} & \multicolumn{1}{c}{$\mathbf{0.8289}$} \\ 
			\bottomrule
		\end{tabular}
		\label{A1}
	\end{table}
	
	\subsection*{Experiments on Large Datasets}
	We chose the REFUGE training set as source domain and ORIGA as target domain and used this setting as a case study to verify the performance of our RDR-Net on large datasets. Two state-of-the-art methods (\emph{i.e.}, $\rho$OSAL and BEAL) are used for comparison, which are both open sources. The results are listed in Table~\ref{A2}. It shows that our RDR-Net achieves superior performance over not only the baseline but also other two competing methods, which demonstrates the effectiveness of RDR-Net on large dataset setting.
	
	\begin{table}[!htb]
		\renewcommand{\thetable}{A\arabic{table}}
		\caption{Performance of two baseline settings, two UDA methods, and our RDR-Net in OD/OC segmentation, when using REFUGE training set as source domain and ORIGA as target domain.}
		\setlength{\tabcolsep}{3pt}
		\centering
		\scriptsize
		\begin{tabular}{m{15pt}<{\centering}m{10pt}<{\centering}m{15pt}<{\centering}m{15pt}<{\centering}}
			\toprule
			\multicolumn{2}{c|}{Method} & \multicolumn{1}{c|}{$Dice_{OD}$} & \multicolumn{1}{c}{$Dice_{OC}$} \\ \hline
			\multicolumn{2}{c|}{No Adapt} & \multicolumn{1}{c|}{$0.9381$} & \multicolumn{1}{c}{$0.7892$}  \\
			\multicolumn{2}{c|}{Upper bound} & \multicolumn{1}{c|}{$0.9761$} & \multicolumn{1}{c}{$0.8982$}  \\ 
			\multicolumn{2}{c|}{$\rho$OSAL} & \multicolumn{1}{c|}{$0.9412$} & \multicolumn{1}{c}{$0.8484$}  \\ 
			\multicolumn{2}{c|}{BEAL} & \multicolumn{1}{c|}{$0.9499$} & \multicolumn{1}{c}{$0.8521$}  \\ 
			\multicolumn{2}{c|}{Ours} & \multicolumn{1}{c|}{$\mathbf{0.9614}$} & \multicolumn{1}{c}{$\mathbf{0.8837}$} \\
			\bottomrule
		\end{tabular}
		\label{A2}
	\end{table}
	
	\subsection*{Experiments on Other Dataset Setting}
	We also chose the model trained on the ORIGA (source) and Drishti-GS/RIM-ONE-r3 (target) as a case study to evaluate our RDR-Net against two open-source competing methods for completeness, and listed the results in Table~\ref{A3}. It reveal that our RDR-Net outperforms other two competing methods, which is consistent with the conclusions we drew in other settings.
	
	\begin{table}[!htb]
		\renewcommand{\thetable}{A\arabic{table}}
		\caption{Performance of two UDA methods and our RDR-Net in OD/OC segmentation, when using ORIGA as source domain and using Drishti-GS and RIM-ONE-r3 as target domains, respectively.}
		\setlength{\tabcolsep}{3pt}
		\centering
		\scriptsize
		\begin{tabular}{m{15pt}<{\centering}m{10pt}<{\centering}m{15pt}<{\centering}m{15pt}<{\centering}m{15pt}<{\centering}m{15pt}<{\centering}}
			\toprule
			\multicolumn{2}{c|}{Method} & \multicolumn{2}{c|}{Drishti-GS} & \multicolumn{2}{c}{RIM-ONE-r3} \\ \cline{3-6} 
			\multicolumn{2}{c|}{} & \multicolumn{1}{c|}{$Dice_{OC}$}  & \multicolumn{1}{c|}{$Dice_{OD}$} & \multicolumn{1}{c|}{$Dice_{OD}$} & \multicolumn{1}{c}{$Dice_{OC}$} \\ \hline
			\multicolumn{2}{c|}{$\rho$OSAL} & \multicolumn{1}{c|}{$0.9718$} & \multicolumn{1}{c|}{$0.8686$}  & \multicolumn{1}{c|}{$0.8885$} & \multicolumn{1}{c}{$0.7754$} \\
			\multicolumn{2}{c|}{BEAL} & \multicolumn{1}{c|}{$0.9714$} & \multicolumn{1}{c|}{$0.8736$}  & \multicolumn{1}{c|}{$0.9159$} & \multicolumn{1}{c}{$0.7919$} \\
			\multicolumn{2}{c|}{Ours} & \multicolumn{1}{c|}{$\mathbf{0.9757}$} & \multicolumn{1}{c|}{$\mathbf{0.8863}$} & \multicolumn{1}{c|}{$\mathbf{0.9387}$} & \multicolumn{1}{c}{$\mathbf{0.8206}$} \\
			\bottomrule
		\end{tabular}
		\label{A3}
	\end{table}
	
	\subsection*{Statistics of Dice Score}
	In Table~\ref{tab6}, we compared our RDR-Net with other three UDA methods using either Drishti-GS or RIM-ONE-r3 as source domain and using other one and ORIGA as target domains, respectively. To verify the stability of RDR-Net, we also provided the mean and standard deviation values of OC Dice, OD Dice, and mean Dice in Table~\ref{A4}. Note that the results of ISFA are obtained by inheritance, which are not listed in Table~\ref{A4} due to the lack of standard deviation. It shows that our RDR-Net has the best mean and standard deviation values on all scenarios, indicating better performance stability of RDR-Net.
	
	\begin{table}[!htb]
		\renewcommand{\thetable}{A\arabic{table}}
		\caption{Performance (mean$\pm$standard deviation) of two UDA methods and our RDR-Net in OD/OC segmentation, when using either Drishti-GS or RIM-ONE-r3 as source domain and using other one and ORIGA as target domains, respectively.}
		\setlength{\tabcolsep}{3pt}
		\centering
		\scriptsize
		\begin{tabular}{m{35pt}<{\centering}|m{40pt}<{\centering}|m{20pt}<{\centering}|m{38pt}<{\centering}|m{38pt}<{\centering}|m{38pt}<{\centering}}
			\toprule
			Source & Target & Method & $Dice_{OD}$ & $Dice_{OC}$ & Mean Dice \\
			\hline
			\multicolumn{1}{c|}{\multirow{10}{*}{Drishti-GS}} & \multirow{5}{*}{RIM-ONE-r3} & $\rho$OSAL & $0.9153\pm0.0327$ & $0.8159\pm0.1682$ & $0.8656\pm0.0909$
			\\
			\multicolumn{1}{c|}{ } & & BEAL & $0.9208\pm0.0349$ & $0.8001\pm0.1200$ & $0.8605\pm0.0663$
			\\
			\multicolumn{1}{c|}{ } & & Ours & $0.9215\pm0.0302$ & $0.8282\pm0.0890$ & $0.8749\pm0.0386$
			\\ \cline{2-6}
			\multicolumn{1}{c|}{ } & \multirow{5}{*}{ORIGA} & $\rho$OSAL & $0.9445\pm0.0709$ & $0.8498\pm0.0950$ & $0.8972\pm0.0607$
			\\
			\multicolumn{1}{c|}{ } & & BEAL & $0.9490\pm0.0266$ & $0.8524\pm0.0879$ & $0.9007\pm0.0445$
			\\ 
			\multicolumn{1}{c|}{ } & & Ours & $0.9684\pm0.0177$ & $0.8834\pm0.0767$ & $0.9259\pm0.0418$ \\	 	
			\hline
			
			\multicolumn{1}{c|}{\multirow{10}{*}{RIM-ONE-r3}} & \multirow{5}{*}{Drishti-GS} & $\rho$OSAL & $0.9096\pm0.0200$ & $0.8303\pm0.1163$ & $0.8700\pm0.0625$ 
			\\
			\multicolumn{1}{c|}{ } & & BEAL & $0.9293\pm0.0188$ & $0.8054\pm0.1133$ & $0.8674\pm0.0598$ 
			\\
			\multicolumn{1}{c|}{ } & & Ours & $0.9434\pm0.0209$ & $0.8430\pm0.1009$ & $0.8932\pm0.0563$\\ \cline{2-6}
			
			\multicolumn{1}{c|}{ } & \multirow{5}{*}{ORIGA} & $\rho$OSAL & $0.9072\pm0.0255$ & $0.8068\pm0.0985$ & $0.8570\pm0.0542$ \\
			\multicolumn{1}{c|}{ } & & BEAL & $0.9369\pm0.0228$ & $0.8378\pm0.1093$ & $0.8874\pm0.0583$
			\\
			\multicolumn{1}{c|}{ } & & Ours & $0.9521\pm0.0385$ & $0.8463\pm0.0809$ & $0.8992\pm0.0501$\\
			\bottomrule
		\end{tabular}			
		\label{A4}
	\end{table}
	
\end{document}